\documentclass[conference]{IEEEtran}
\IEEEoverridecommandlockouts
\usepackage{cite}
\usepackage{amsmath,amssymb,amsfonts}
\usepackage[ruled,linesnumbered]{algorithm2e}
\usepackage{graphicx}
\usepackage{textcomp}
\usepackage{xcolor}
\usepackage{booktabs}
\usepackage{multirow}
\usepackage{subfig}
\def\BibTeX{{\rm B\kern-.05em{\sc i\kern-.025em b}\kern-.08em
    T\kern-.1667em\lower.7ex\hbox{E}\kern-.125emX}}
\begin{document}

\title{Adaptive CPU Resource Allocation for Emulator in Kernel-based Virtual Machine
}

\author{\IEEEauthorblockN{Yecheng Yang, Pu Pang, Jiawen Wang, Quan Chen, Minyi Guo}

\IEEEauthorblockA{\textit{Department of Computer Science and Engineering, Shanghai Jiao Tong University, China}}
\IEEEauthorblockA{\{yangyecheng, avengerispp, wangjiawen0606\}@sjtu.edu.cn, \{chen-quan, guo-my\}@cs.sjtu.edu.cn}}

\maketitle

\begin{abstract}
The technologies of heterogeneous multi-core architectures, co-location, and virtualization can be used to reduce server power consumption and improve system utilization, which are three important technologies for data centers. This article explores the scheduling strategy of Emulator threads within virtual machine processes in a scenario of co-location of multiple virtual machines on heterogeneous multi-core architectures. In this co-location scenario, the scheduling strategy for Emulator threads significantly affects the performance of virtual machines. This article focuses on this thread for the first time in the relevant field. This article found that the scheduling latency metric can well indicate the running status of the vCPU threads and Emulator threads in the virtualization environment, and applied this metric to the design of the scheduling strategy. This article designed an Emulator thread scheduler based on heuristic rules, which, in coordination with the host operating system's scheduler, dynamically adjusts the scheduling scope of Emulator threads to improve the overall performance of virtual machines. The article found that in real application scenarios, the scheduler effectively improved the performance of applications within virtual machines, with a maximum performance improvement of 40.7\%.
\end{abstract}


\section{Introduction}

As data centers continue to evolve, the exploration and application of heterogeneous multi-core architecture, co-location, and virtualization technologies are becoming increasingly prevalent. These three technologies each have their unique roles and significant implications for data centers. This paper aims to integrate these three technologies and delve into the domain of co-location virtual machine deployment within a heterogeneous multi-core architecture environment.

The heterogeneous multi-core architecture refers to a technology that utilizes heterogeneous multi-core central processing units (CPUs). The cores within this architecture share the same instruction set but differ in their computational capabilities. Big cores are characterized by their higher computational power, albeit at the cost of increased power consumption and energy usage. Small cores, on the other hand, offer lower computational capabilities but are more power-efficient, boasting a higher energy-to-performance ratio. The key to heterogeneous multi-core technology is the mixed utilization of these two distinct types of computing cores, leveraging the unique advantages of each. When only big cores are used, the system exhibits strong overall computational capabilities but consumes more power. Conversely, with only small cores, the system becomes more power-efficient. However, when faced with performance-intensive, time-sensitive tasks, it may struggle to meet their performance and Quality of Service (QoS) requirements.

Co-location technology involves deploying a certain number of applications on a single physical server. In data centers, a multi-tenant approach is often used to mix and deploy tasks from multiple users on a single physical node, thereby increasing server utilization and aiding data centers in improving cost-effectiveness \cite{barroso2019datacenter}. Some studies \cite{kaplan2008revolutionizing,vasan2010worth,reiss2012heterogeneity,delimitrou2014quasar,carvalho2014long,barroso2019datacenter} have shown that the average utilization of most data centers is quite low, fluctuating between 10\% and 50\%. This low utilization results in a significant waste of computational resources, and it has been demonstrated that lower utilization rates can decrease processor energy efficiency \cite{barroso2007case}. Research findings \cite{lo2015heracles,chen2019parties,patel2020clite} indicate that by employing co-location technology to mix and deploy certain tasks on a single physical machine, server utilization can be effectively improved.

Virtualization technology refers to the simulation of one or multiple virtual machines on a physical host using a combination of hardware and software techniques. Virtual machines created through virtualization technology can quickly perform management tasks such as creation, destruction, migration, and scaling. Moreover, these virtual machines can effectively isolate themselves from one another, allowing multiple virtual machines to be mixed and deployed on the same physical node to realize multi-tenancy characteristics and enhance server utilization \cite{kwon2011virtualizing}. Due to its numerous advantages, virtualization technology is highly favored within data centers.

Virtualization technology can be implemented in various ways, with Kernel-based Virtual Machine (KVM) being a widely used virtualization technology in the Linux ecosystem. KVM is presented as a kernel module within the kernel and a QEMU (Quick Emulator) process in user space. In essence, a KVM virtual machine appears as a regular process in user space, enabling administrators to manage it using typical process management methods. The QEMU process encompasses two types of threads: virtual CPU (vCPU) threads and virtualization (Emulator) threads. vCPU threads serve to virtualize the processor, executing binary code just like a real processor would for the system's required tasks. Emulator threads take on the role of executing input/output (I/O) operations on behalf of vCPU threads to prevent vCPU threads from becoming blocked. Both types of threads are scheduled by the system's scheduler, much like regular threads within a system.

Building upon the three aforementioned technologies – heterogeneous multi-core architecture, co-location, and virtualization – this paper aims to explore the co-location of virtual machines in a heterogeneous multi-core architecture environment. In this exploration, particular attention will be given to the scheduling methods for both vCPU and Emulator threads within the context of co-location. More specifically, this paper intends to investigate the deployment of multiple KVM virtual machines on a single physical node within a heterogeneous multi-core architecture environment. Such an environment features both powerful, energy-intensive big cores and less powerful, energy-efficient small cores. Throughout this co-location process, the scheduling strategies for vCPU and Emulator threads will be taken into account, with the goal of reducing interference between these two thread types and enhancing application performance within virtual machines.



In the scenario presented in this paper, the Emulator thread is a noteworthy component, and its scheduling and deployment strategy can have a substantial impact on virtual machine performance in certain scenarios. Conversely, when handled appropriately in this scenario, the scheduling and deployment of Emulator threads can lead to performance improvements within virtual machines. Therefore, this research scenario is worth exploring, and the scheduling strategies for Emulator threads merit thorough study.

The research in this paper encompasses an analysis of the impact of Emulator thread utilization, optimal core binding, and its scheduling deployment strategy on application performance within a virtual machine while running different applications. Several key findings include indicators of Emulator thread and vCPU thread operational states, providing valuable metrics for scheduling. An intelligent rule-based scheduler has been devised and tested on a physical machine to validate its scheduling effectiveness, comparing it to a baseline.

In summary, the specific contributions of this paper are as follows:

\begin{enumerate}
    \item An examination of the disparities in Emulator thread utilization and optimal core binding when running various applications within a virtual machine, along with an analysis of the influence of Emulator scheduling and deployment strategies on application performance. 
    
    \item The proposal of the "Run Delay" metric as a critical indicator for scheduling Emulator threads, effectively reflecting the operational states of both Emulator and vCPU threads. Balancing the Run Delay of Emulator and vCPU threads is essential for improving virtual machine performance.
    
    \item The design of a heuristic-based scheduler, built upon the Linux Completely Fair Scheduler (CFS), which controls the scheduling scope of Emulator threads based on the virtual machine's status. 
    
    \item Verification of the scheduler's effectiveness, showcasing its ability to enhance virtual machine performance in real-world scenarios. Compared to the baseline, it achieved a remarkable performance improvement of up to 40.7\%. 
\end{enumerate}

\section{Related Work}

This section introduces relevant work in the domains of heterogeneous multi-core architecture, co-location, and virtualization.

\subsection{Relevant Work in the Heterogeneous Multi-core Architecture Context}

In this section, the research scenario revolves around a co-location of time-sensitive tasks with Quality of Service (QoS) requirements and several batch processing tasks within the context of heterogeneous multi-core architecture. Petrucci et al. \cite{petrucci2015octopus} introduced Octopus-Man, a work that employs a heuristic approach and an automatic control method to dynamically schedule tasks on either several big cores or several small cores. Nishtala et al. \cite{nishtala2017hipster} proposed Hipster, which employs reinforcement learning to select an appropriate core configuration for time-sensitive tasks based on their current QoS status. 


These works inspire the present study in the consideration of heuristic or machine learning methods for designing scheduling strategies for applications running on a single bare machine, without considering virtualized environments.

\subsection{Relevant Work in Co-Location}

The research context in this section is based on the deployment of one or multiple time-sensitive tasks with QoS requirements alongside various batch processing tasks in a conventional homogeneous core environment. Chen et al. \cite{chen2019parties} introduced PARTIES, a system designed to run multiple time-sensitive tasks on a single system while guaranteeing their QoS. Nishtala et al. \cite{nishtala2020twig} presented Twig, which differs from PARTIES by utilizing reinforcement learning to manage resource allocation for multiple time-sensitive tasks. These works consider the co-location of multiple time-sensitive tasks in a single-machine environment, they do not take into account the specific characteristics of the heterogeneous multi-core architecture and virtualized environments.

\subsection{Virtualization-Related Work}

Some of the research in this category focuses on the co-location of multiple virtual machines in a single homogeneous core environment. Xu et al. \cite{xu2013vturbo} introduced vTurbo, emphasizing that a significant reason for degraded I/O performance in virtual machines is interrupt processing latency. To address this issue, vTurbo reduces the scheduling time slice of one or more cores and assigns interrupt handling tasks within the virtual machine to these cores. The insights from this work suggest that Emulator threads should be given dedicated access to one or more specific physical cores. These works are concerned with I/O task management within virtual machines, particularly relevant to Emulator threads in the KVM environment. However, they primarily focus on vCPU threads and do not directly address the scheduling of Emulator threads.

Another subset of research within this category explores the co-location of multiple virtual machines in a heterogeneous multi-core architecture environment. Kwon et al. \cite{kwon2011virtualizing} proposed a scheduling method for virtual machines in a heterogeneous multi-core architecture setting. This work assigns acceleration ratio to each virtual machines. Virtual machines with a higher acceleration ratio are given priority access to big cores, aiming to enhance overall system throughput and performance. However, the main focus of this work is on vCPU threads.

\section{Motivation}

\subsection{Analysis of Emulator Thread Utilization}

This section delves into the utilization of Emulator threads while running various applications within the virtual machine. It was observed that the utilization of Emulator threads varies significantly depending on the specific application being run. Notably, when running Memcached and Nginx, the Emulator thread utilization within the virtual machine was remarkably high, nearly reaching half of the entire virtual machine process's utilization. Researchers have noted, as per Liu et al. (2019) \cite{liu2019vcpu}, that in Xen virtualization environments, virtual machines with frequent I/O operations can consume up to 90\% of CPU resources on the I/O processing domain. In this section, a similar phenomenon was observed in the context of the KVM environment.

\begin{figure}[!htp]
\begin{minipage}[t]{0.48\textwidth}
  \vspace{0pt}
  \centering
  \centerline{\includegraphics[width=8cm]{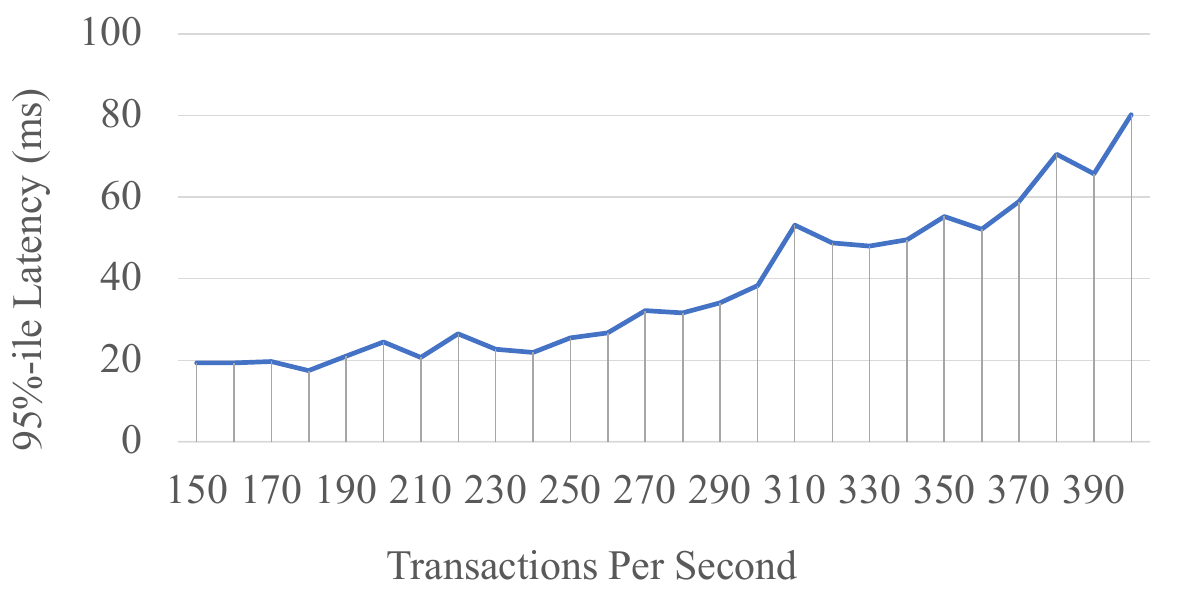}}
  \caption{The variation of MySQL's P95 latency with varying workloads}
  \label{fig:mysql-p95}
\end{minipage}\hfill
\begin{minipage}[t]{0.48\textwidth}
  \vspace{0pt}
  \centering
  \centerline{\includegraphics[width=8cm]{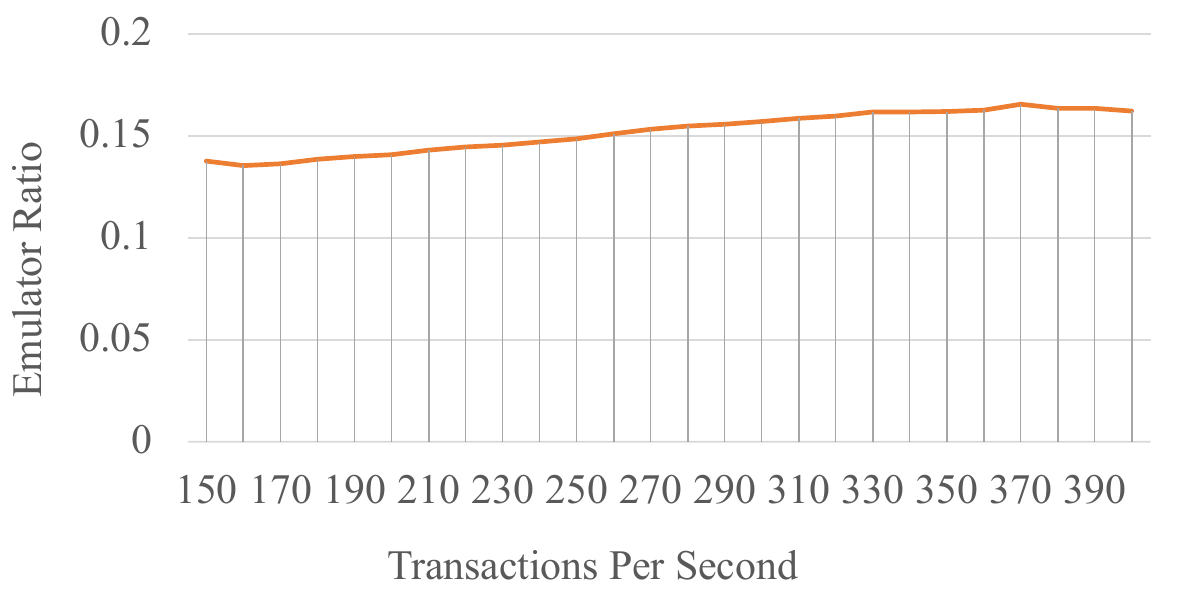}}
  \caption{The variation of MySQL's Emulator Ratio with varying workloads}
  \label{fig:mysql-ratio}
\end{minipage}
\end{figure}


Figures \ref{fig:mysql-p95} and \ref{fig:mysql-ratio} depict the tail latency and Emulator Ratio values while running the MySQL application within the virtual machine under varying workloads. The horizontal axis in both figures represents Transactions Per Second (TPS), where higher TPS values indicate heavier MySQL workloads. The vertical axes represent P95 tail latency and Emulator Ratio. Emulator Ratio is defined as:

\begin{align}
\begin{split}
&Emulator\ Ratio = \\
&\quad\frac{cpuUsage(Emulator)}{cpuUsage(Emulator) + cpuUsage(vCPU)}.
\end{split}
\end{align}

Here, cpuUsage represents the CPU runtime of a specific thread, which can be obtained through operating system-level interfaces. The Emulator Ratio value falls within the range of 0 to 1. A value closer to 1 indicates a higher proportion of CPU utilization by the Emulator thread. It can be observed that with an increase in MySQL workload, both MySQL's tail latency and Emulator Ratio gradually rise.

\begin{figure*}[!htp]
\centering
\subfloat[The variation of Memcached's P95 latency with varying workloads\label{fig:memcached-p95}]{\includegraphics[width=4cm]{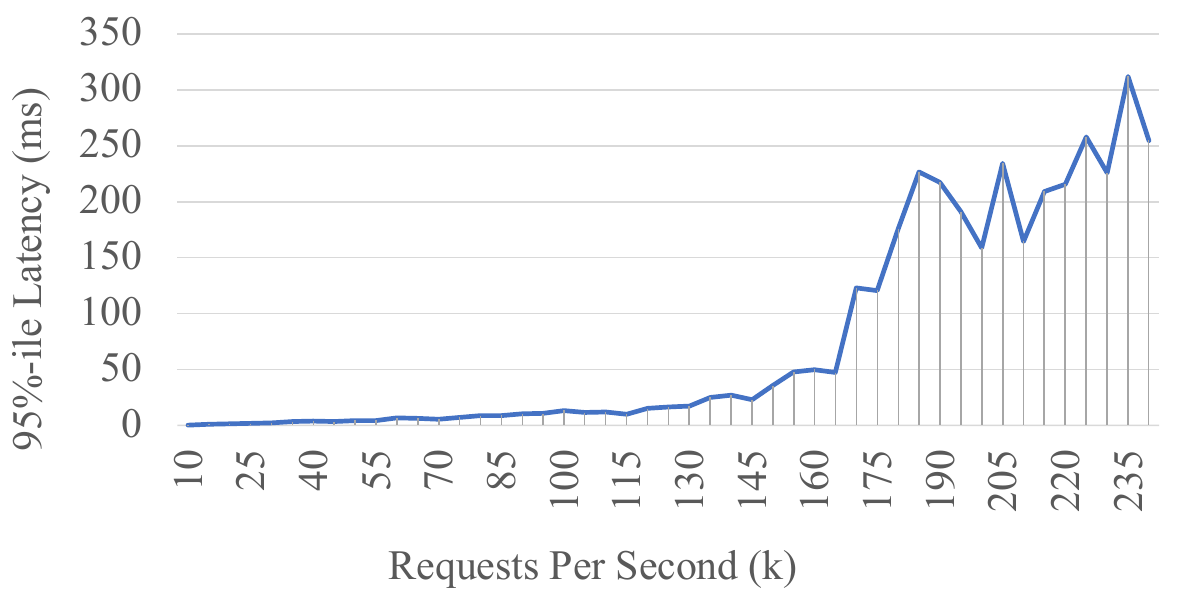}}
\hfil
\subfloat[The variation of Memcached's Emulator Ratio with varying workloads\label{fig:memcached-ratio}]{\includegraphics[width=4cm]{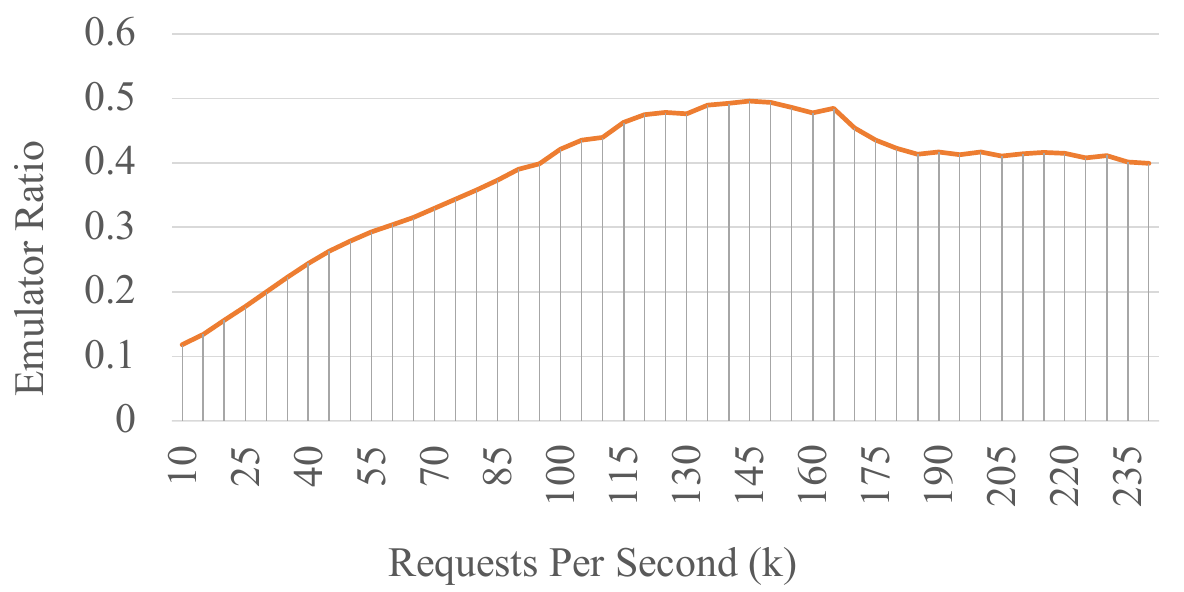}}
\hfil
\subfloat[The variation of Nginx's Emulator Ratio with varying workloads\label{fig:nginx-p95}]{\includegraphics[width=4cm]{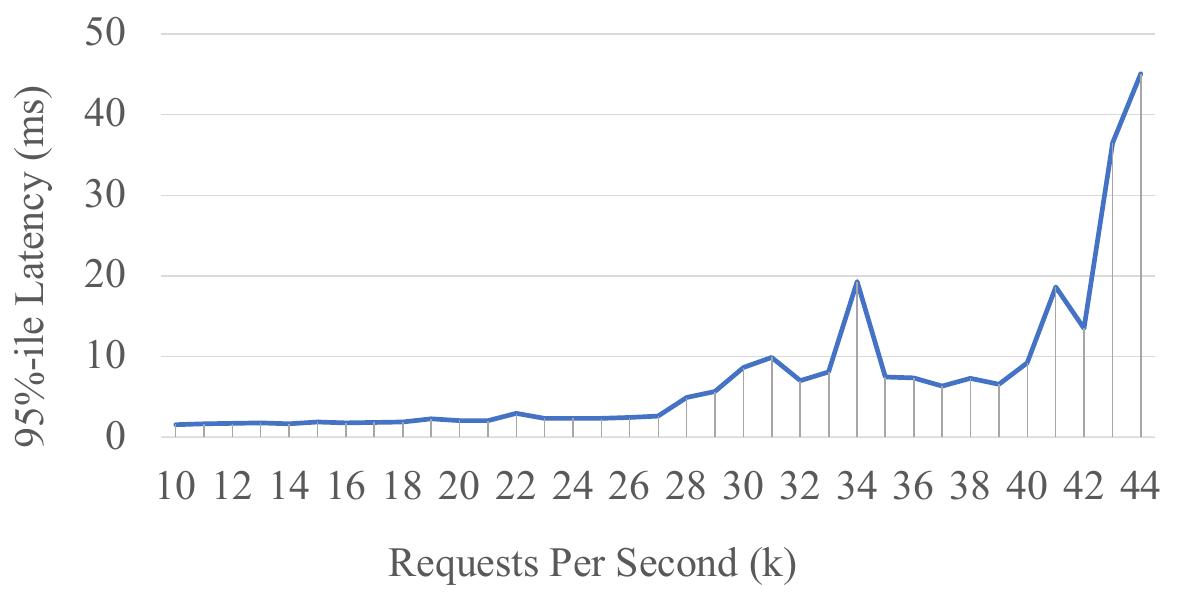}}
\hfil
\subfloat[The variation of Nginx's Emulator Ratio with varying workloads\label{fig:nginx-ratio}]{\includegraphics[width=4cm]{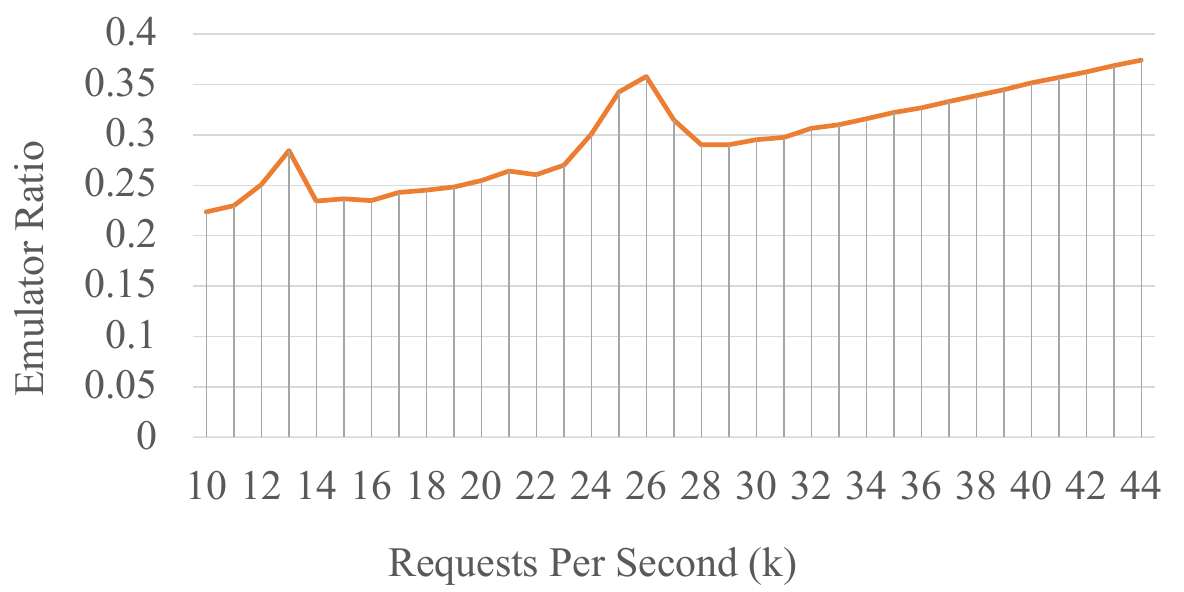}}
\hfil
\\
\subfloat[The variation of Xapian's Emulator Ratio with varying workloads\label{fig:xapian-p95}]{\includegraphics[width=4cm]{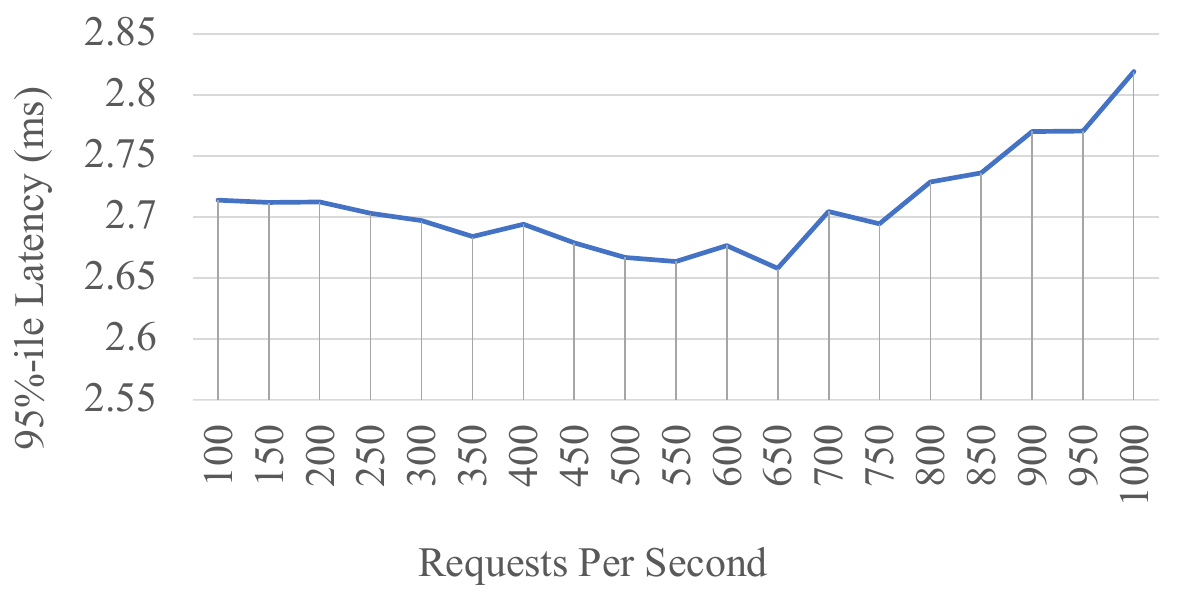}}
\hfil
\subfloat[The variation of Xapian's Emulator Ratio with varying workloads\label{fig:xapian-ratio}]{\includegraphics[width=4cm]{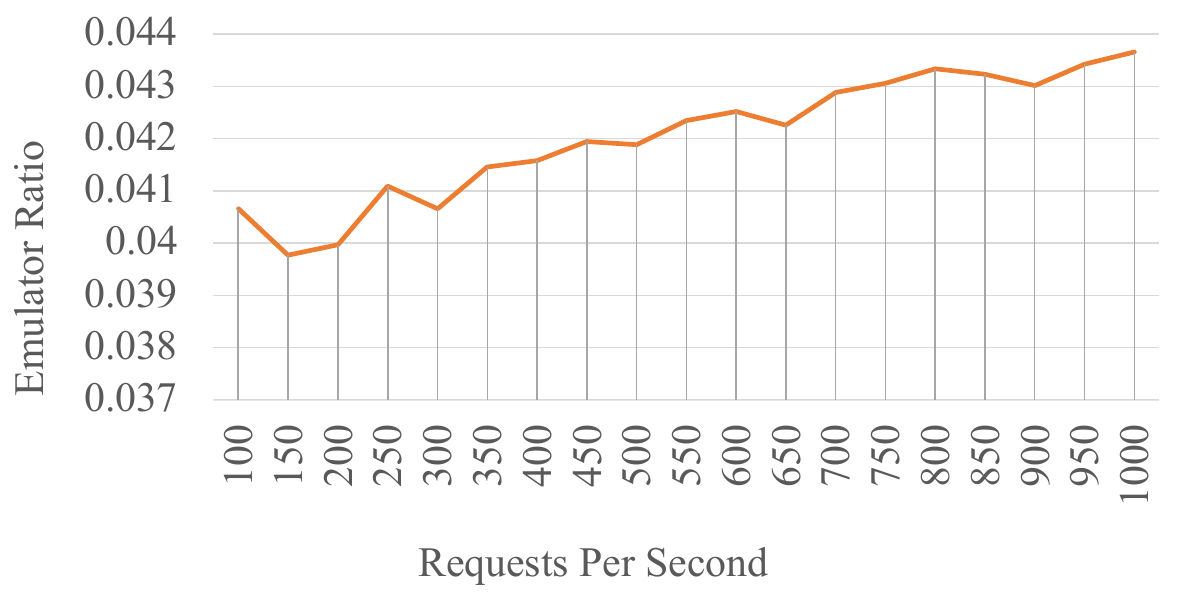}}
\hfil
\subfloat[The variation of PostgreSQL's Emulator Ratio with varying workloads\label{fig:pg-p95}]{\includegraphics[width=4cm]{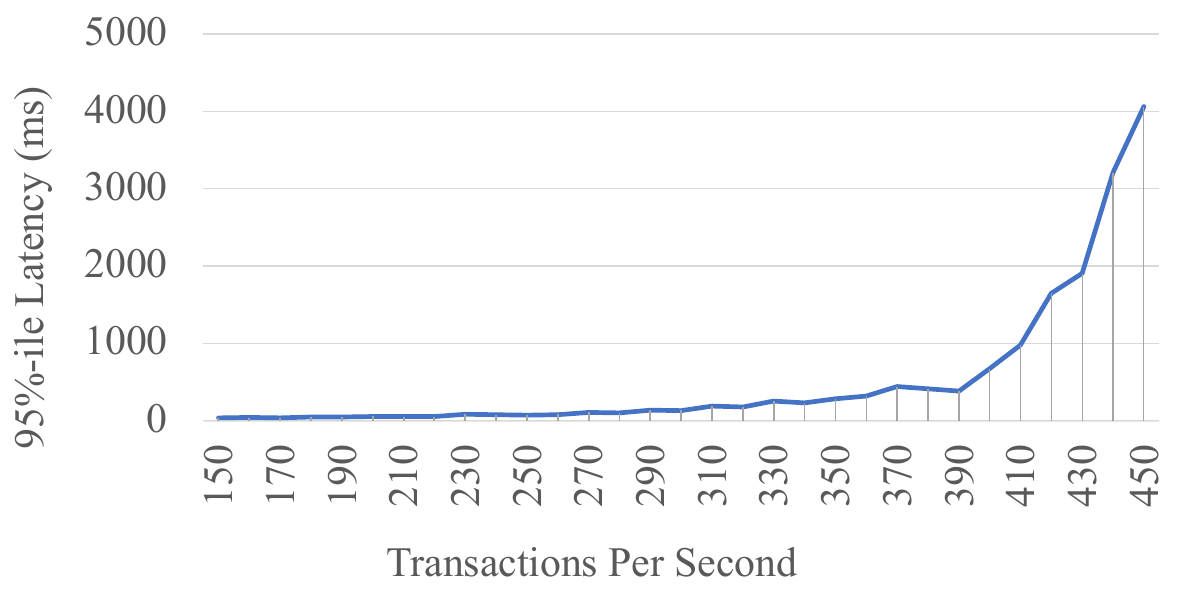}}
\hfil
\subfloat[The variation of PostgreSQL's Emulator Ratio with varying workloads\label{fig:pg-ratio}]{\includegraphics[width=4cm]{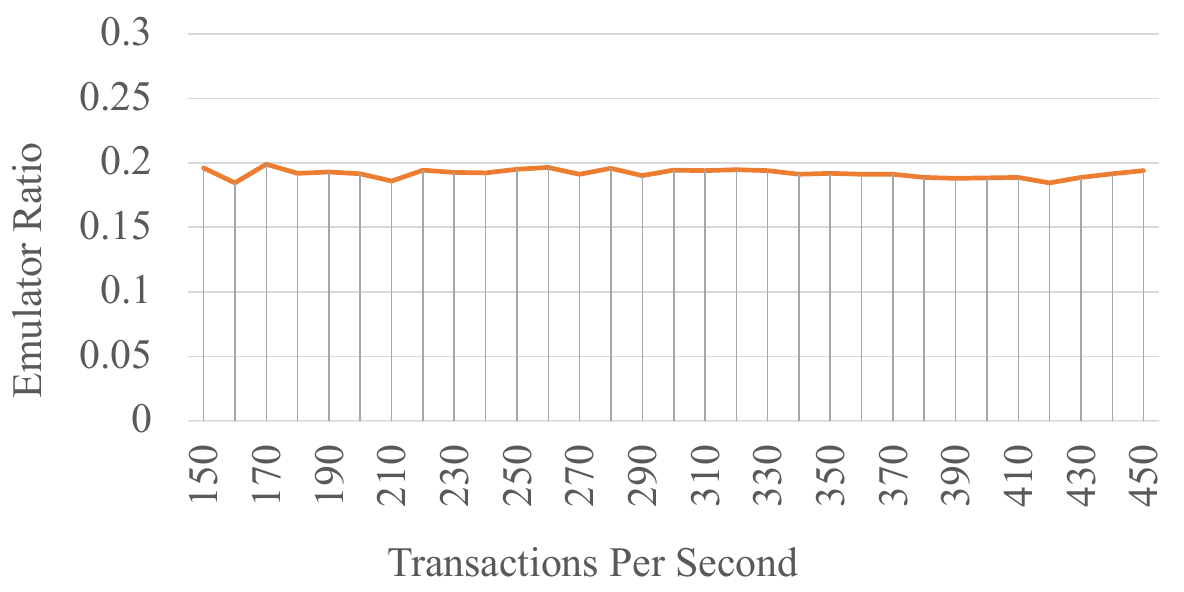}}
\hfil

\caption{The variation of P95 latency and Emulator Ratio with varying workloads}
\label{fig:p95-and-ratio}
\end{figure*}

Similar trends are observed in other applications. Figures \ref{fig:memcached-p95}, \ref{fig:memcached-ratio}, \ref{fig:nginx-p95}, \ref{fig:nginx-ratio}, \ref{fig:xapian-p95}, \ref{fig:xapian-ratio}, \ref{fig:pg-p95}, and \ref{fig:pg-ratio} illustrate the situations corresponding to Memcached, Nginx, Xapian, and PostgreSQL. The variations in tail latency and Emulator Ratio under high workload conditions differ slightly among these applications. With the exception of PostgreSQL, where the Emulator Ratio remains relatively stable, the overall trend for both values is an upward trajectory with increasing workloads.

The Emulator Ratio value reflects the proportion of CPU usage by Emulator threads relative to all virtual machine threads. The results highlight that the Emulator's share of CPU utilization is significant, underscoring the need for scheduling policies tailored to Emulator threads. Furthermore, the variations in Emulator Ratio values and their rate of change under different workloads differ across applications. The diverse characteristics among applications pose a substantial challenge in selecting suitable scheduling methods. It is worth noting that despite the notable Emulator Ratio values, they remain below 0.5 for all applications, indicating that vCPU threads still dominate the majority of CPU usage time. This underscores the importance of considering vCPU threads in the design of scheduling strategies presented in this study.

\subsection{Analysis of Emulator Thread Core Binding}

This section explores the impact of different core bindings for Emulator threads on virtual machine performance when running various applications within the virtual machine. Binding, in this context, refers to the process of utilizing operating system-level interfaces to restrict Emulator thread scheduling to a specific set of physical CPU cores. The specific timing and core selection for scheduling are determined by the Linux system's default scheduler, the Completely Fair Scheduler (CFS).

\begin{figure}[!htp]
  \centering
  \includegraphics[width=8cm]{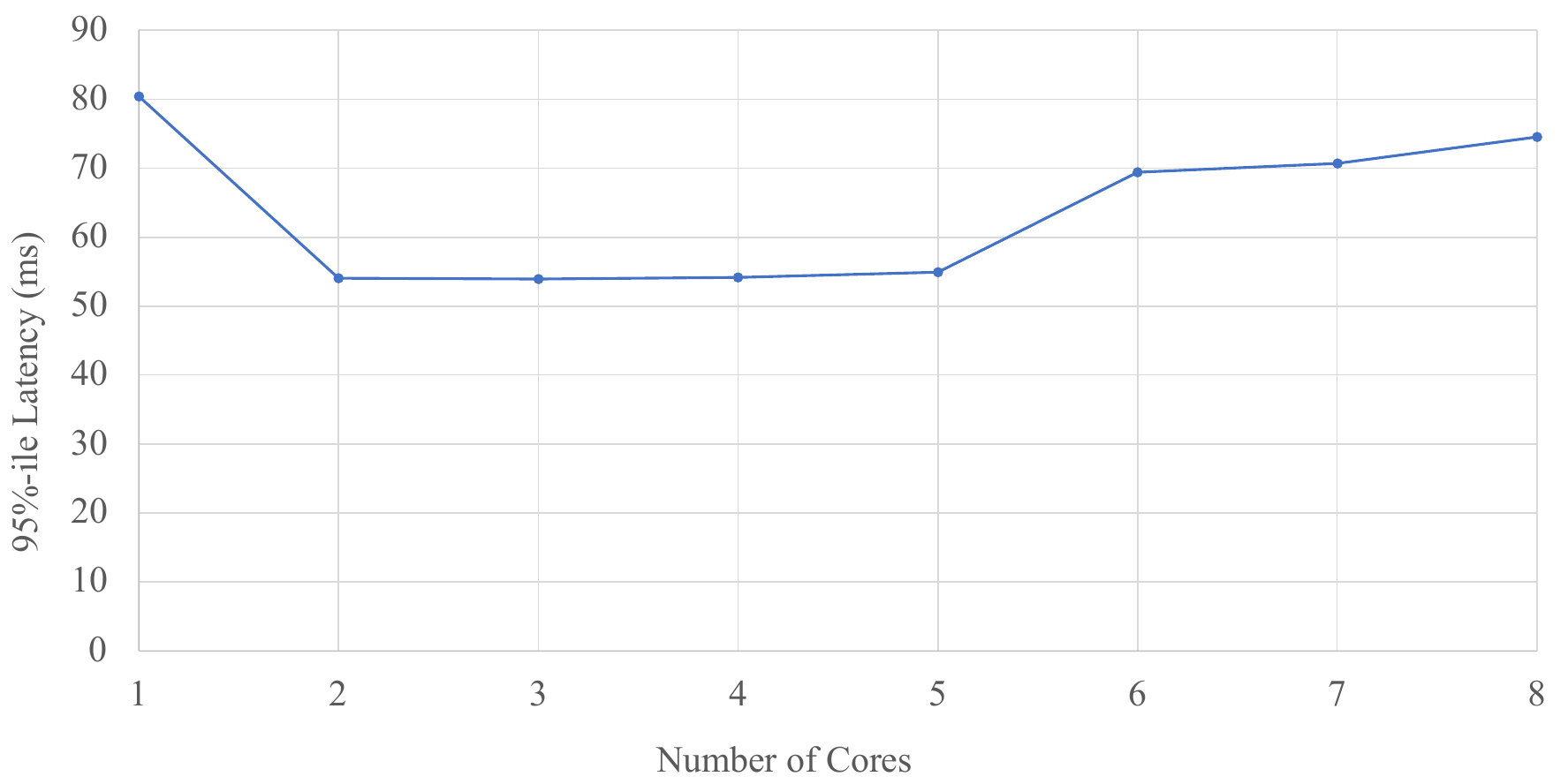}
  \caption{Impact of Different Emulator Thread Core Bindings on MySQL's P95 Latency}
  \label{fig:mysql-p95-cores}
\end{figure}

It is found that the optimal number of bound cores varies among different applications. Figure \ref{fig:mysql-p95-cores} illustrates the situation for MySQL. When the number of bound cores is minimal, MySQL's performance is suboptimal. Slightly increasing the number of bound cores results in performance improvements, but further increases lead to performance degradation. For MySQL, the optimal number of bound cores is determined to be 3.

\begin{figure*}[!htp]
\centering
\subfloat[Impact of Different Emulator Thread Core Bindings on Memcached's P95 Latency\label{fig:memcached-p95-cores}]{\includegraphics[width=4cm]{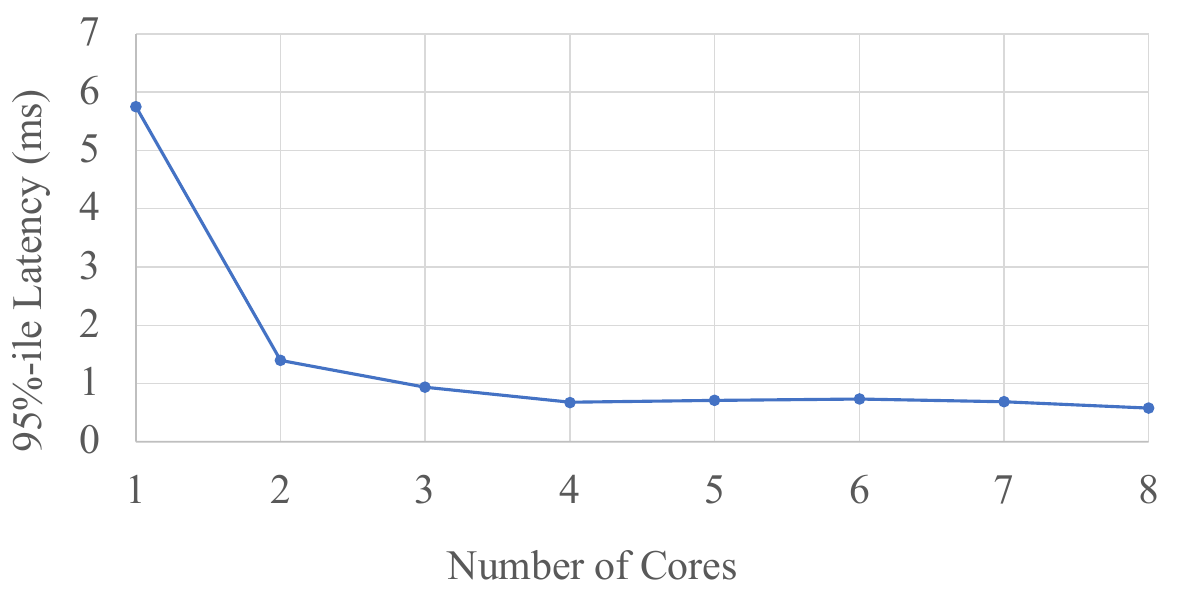}}
\hfil
\subfloat[Impact of Different Emulator Thread Core Bindings on Nginx's P95 Latency\label{fig:nginx-p95-cores}]{\includegraphics[width=4cm]{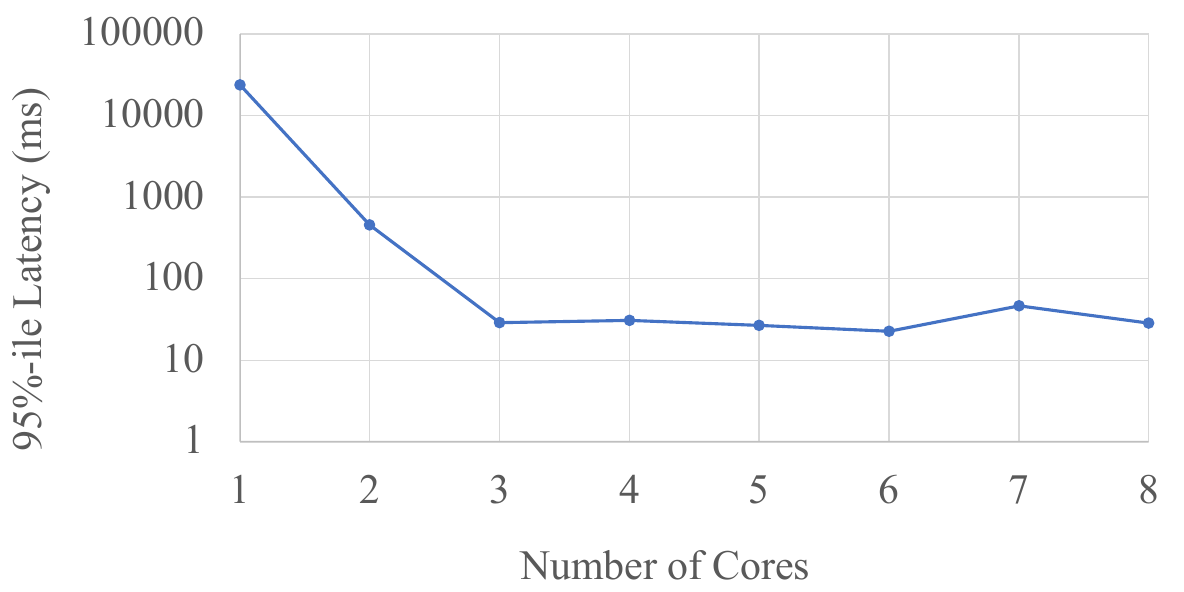}}
\hfil
\subfloat[Impact of Different Emulator Thread Core Bindings on Xapian's P95 Latency\label{fig:xapian-p95-cores}]{\includegraphics[width=4cm]{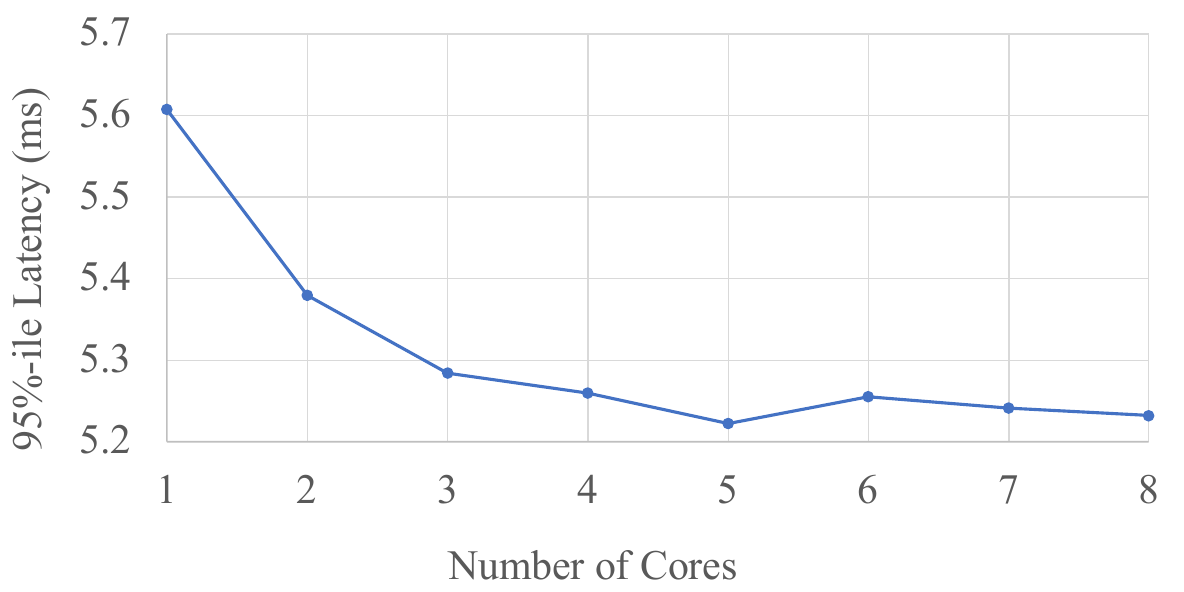}}
\hfil
\subfloat[Impact of Different Emulator Thread Core Bindings on PostgreSQL's P95 Latency\label{fig:pg-p95-cores}]{\includegraphics[width=4cm]{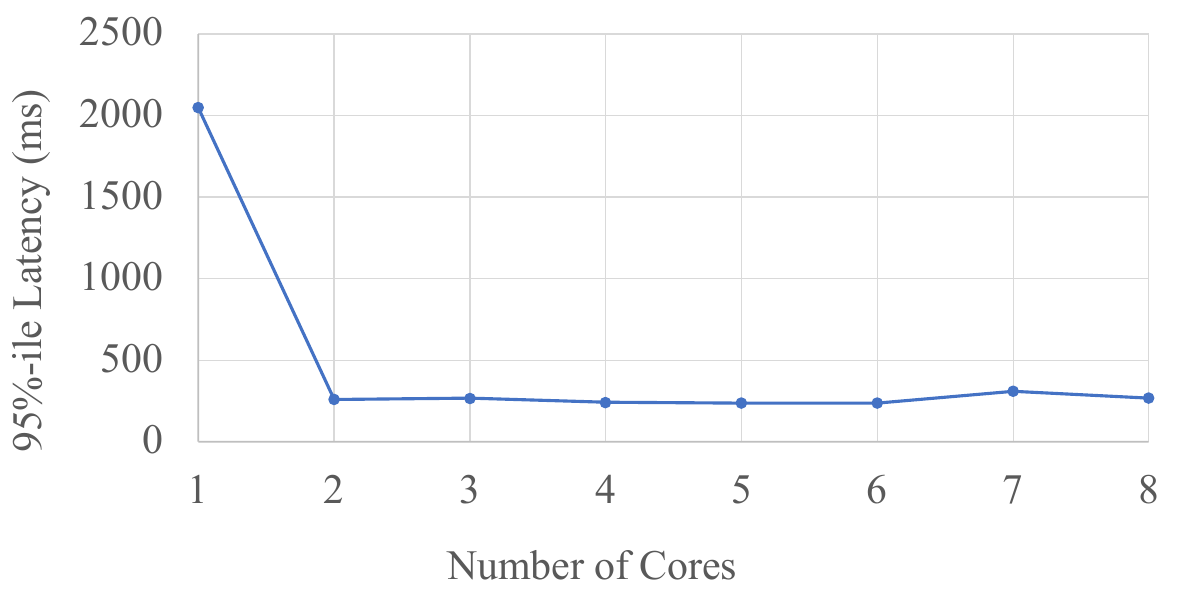}}
\hfil

\caption{Impact of Different Emulator Thread Core Bindings on application's P95 Latency}
\label{fig:p95-and-cores}
\end{figure*}

Figures \ref{fig:memcached-p95-cores}, \ref{fig:nginx-p95-cores}, \ref{fig:xapian-p95-cores}, and \ref{fig:pg-p95-cores} present the scenarios for Memcached, Nginx, Xapian, and PostgreSQL. In contrast to MySQL, these applications do not exhibit significant performance degradation when a larger number of cores are bound. Among these applications, the optimal core binding numbers are as follows: 8 for Memcached, 6 for Nginx, 5 for Xapian, and 6 for PostgreSQL. The fact that different applications require varying optimal core bindings presents a challenge in determining the most suitable core binding configuration for each application.

\subsection{Analysis of Static Scheduling Strategies for Emulator Threads in Co-location}

This section analyzes the impact of statically adjusting Emulator thread scheduling strategies on the performance of virtual machines in a co-location environment. Specifically, this section begins with a theoretical analysis of the total number of Emulator thread scheduling strategies available for hybrid-deployed virtual machines and then explores the influence of these strategies on virtual machine performance.

\subsubsection{Analysis of the Static Scheduling Strategy Space}

Assuming there are $n$ small cores and $m$ large cores on a physical machine, and a total of $x$ virtual machines are deployed. The question is how many Emulator scheduling strategies exist. A scheduling strategy is defined as the allocation of Emulator threads to the cores for the $x$ virtual machines. Formally, let set $N$ be the set of small cores and set $M$ be the set of large cores. A scheduling strategy can be represented as an ordered set $X' = \left\{Y'_i|Y'_i \subseteq N \cup M, 1\le i \le x\right\}$, where $Y'_i$ is the scheduling strategy for each virtual machine.

We find that there are equivalence relations between scheduling strategies. Through mathematical calculations, it can be determined that there $t$ fundamentally different scheduling strategies. $t$ can be determined as follows:

\begin{equation}
t = C^{n}_{2^x+n-1} \times C^{m}_{2^x+m-1}. 
\end{equation}

It's evident that $t$ rapidly increases with the values of $n$, $m$, and $x$. A significant challenge in this study lies in how to select the best scheduling strategy from this vast pool of scheduling strategies.

\subsubsection{Analysis of the Impact of Static Scheduling Strategies on Application Performance}
This subsection analyzes the impact of 220 different scheduling strategies on the performance of applications within virtual machines. To limit the enumeration space, parameters are set to satisfy $n=4$, $m=4$, $x=2$, and $|Y_i| \in \left\{1,2,4,8\right\}$ for all $i \in \left\{1, 2\right\}$. Two virtual machines are deployed, each running MySQL and Memcached, resulting in a total of 220 fundamentally distinct scheduling strategies under these conditions.

The 220 scheduling strategies have a significant impact on application performance within the virtual machines. The best scheduling strategies can improve MySQL performance by 75.4\%, Memcached performance by 64.2\%, compared to the worst scheduling strategies. Compared to the baseline scenario (without specific scheduling, as detailed in subsequent chapters), MySQL performance can be improved by 50.4\%, and Memcached can be improved by 38.3\%. This demonstrates that good scheduling strategies can greatly enhance application performance, while poor scheduling strategies can significantly degrade it. A major challenge of this study is to identify good strategies and avoid poor ones.

Additionally, out of the 220 effective strategies, many tend to favor the use of small cores. The best strategy involves both virtual machines sharing the same two small cores. 



\subsection{Analysis of Emulator Thread Scheduling Metrics in Co-location Environment}\label{sec:rundelay-analysis}

This section analyzes the scheduling metrics of Emulator threads, specifically focusing on "Run Delay",  which refers to the amount of time a thread waits in the CPU scheduling queue. When this value becomes significant, it indicates that the thread has spent a considerable amount of time waiting on the queue and has not been able to access CPU time. A high Run Delay for a thread can negatively impact its performance.

\begin{figure}[!htp]
  \centering
  \includegraphics[width=8cm]{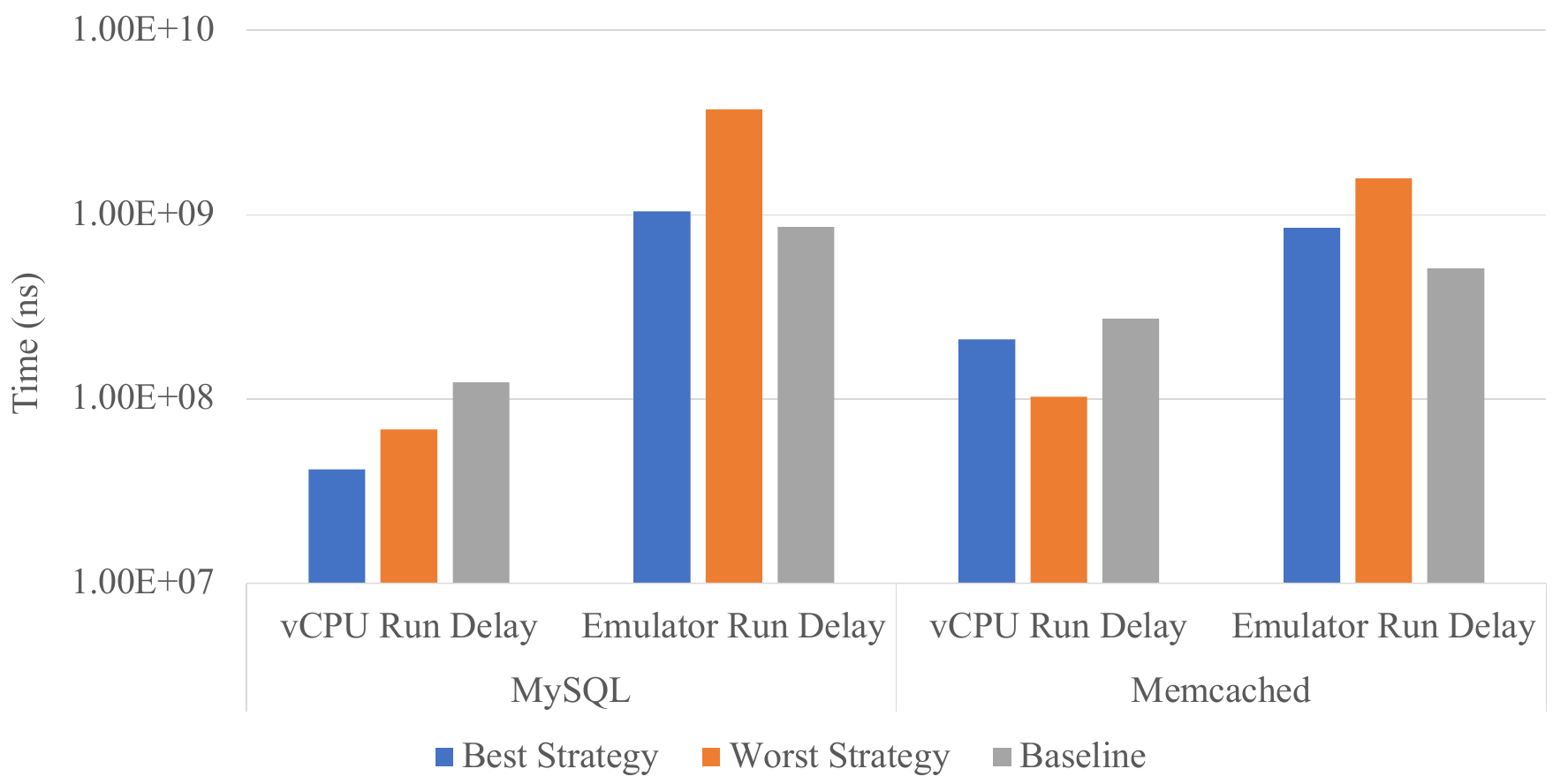}
  \caption{Scheduling Metrics for Different Scheduling Strategies in a Co-location of MySQL and Memcached}
  \label{fig:mysql-memcached-rundelay}
\end{figure}

Taking the example of a co-location of MySQL and Memcached, Figure \ref{fig:mysql-memcached-rundelay} illustrates the scheduling metrics for different scheduling strategies. It is evident that the worst scheduling strategy, while reducing vCPU scheduling delay compared to the baseline scenario, increases Emulator scheduling delay, resulting in poorer performance. Conversely, the baseline scenario, compared to the best scheduling strategy, reduces Emulator scheduling delay but increases vCPU scheduling delay, leading to suboptimal performance. To maximize virtual machine performance, it is necessary to balance the scheduling delays of vCPU and Emulator threads. A significant challenge in this study is finding the right balance between these two scheduling delays.

\begin{figure}[!htp]
  \centering
  \includegraphics[width=8cm]{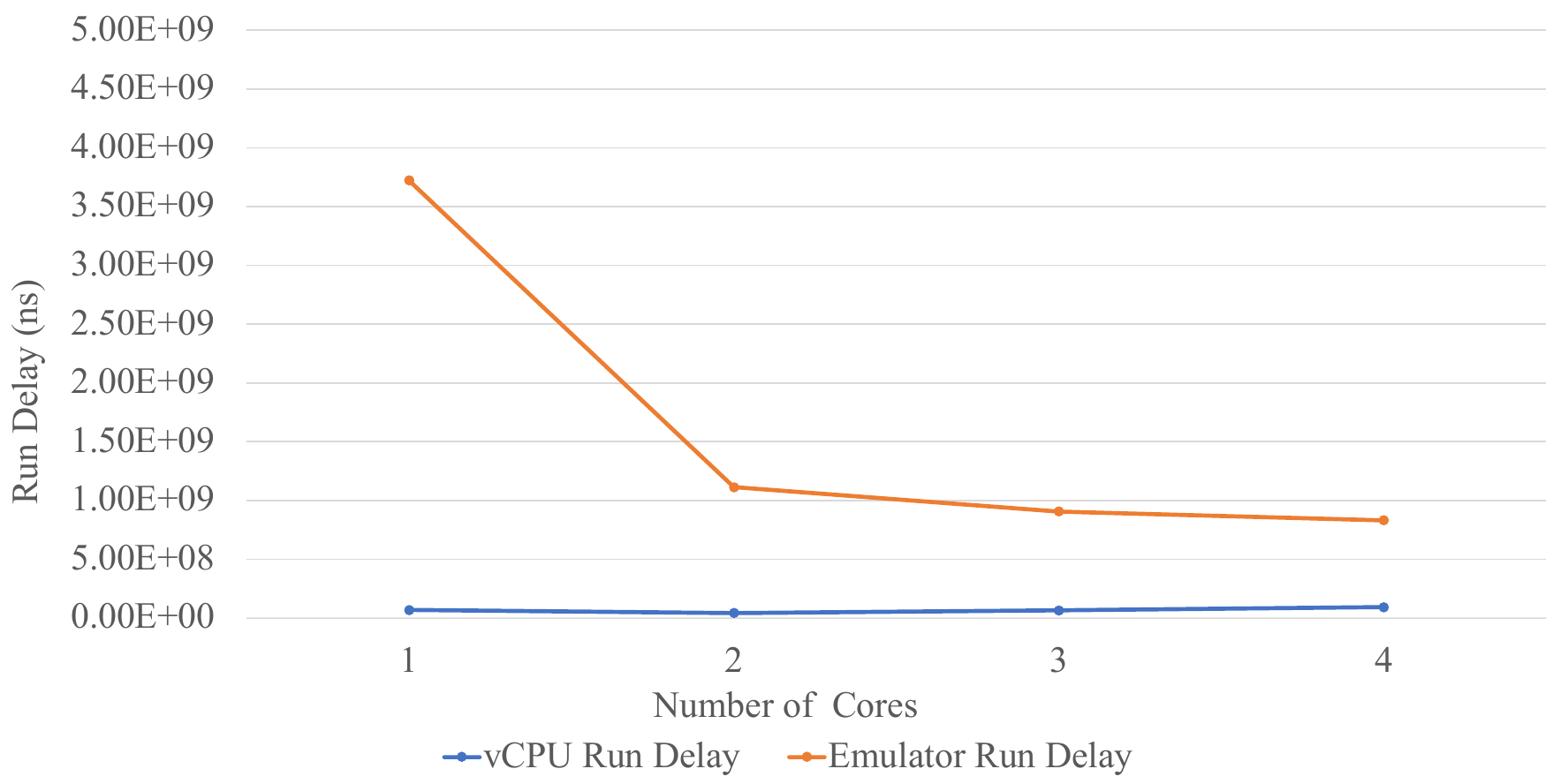}
  \caption{Impact of Changing the Number of Emulator Threads Bound to MySQL Virtual Machine Cores on Scheduling Delays in a Co-location of MySQL and Memcached}
  \label{fig:mysql-memcached-rundelay-cores}
\end{figure}

To analyze the balance point for scheduling delays, while keeping the binding method of Memcached fixed, the number of Emulator threads bound to MySQL virtual machine cores is increased. Figure \ref{fig:mysql-memcached-rundelay-cores} illustrates the change in scheduling delays for MySQL virtual machine vCPU and Emulator threads. It can be observed that as the number of Emulator threads bound to MySQL virtual machine cores increases, the scheduling delay for Emulator threads rapidly decreases initially, followed by a slower decrease, while the scheduling delay for vCPU threads shows an upward trend. To balance the scheduling delays for both threads, it is essential to minimize the scheduling delay for both vCPU and Emulator threads. The point at which the decrease in Emulator thread scheduling delay begins to slow down is an appropriate balance point. Experiments have shown that this point is indeed close to the optimal scheduling strategy.

\begin{table}[!hpt]
  \caption{CPU Utilization Disparity for Different Scheduling Strategies in a Co-location of MySQL and Memcached}
  \label{tab:mysql-memcached-cpu-max-min}
  \centering
  \begin{tabular}{@{}ccc@{}} \toprule
    Scheduling Strategy & CPU Utilization Disparity\\ \midrule
    Best Strategy & 0.128 \\
    Worst Strategy & 0.434 \\
    Baseline Strategy & 0.050 \\ \bottomrule
  \end{tabular}
\end{table}

Additionally, the balance of physical core utilization in virtual machines reflects the overall system condition. A more significant imbalance in physical core utilization indicates that some physical cores are experiencing higher utilization. This might signify that several virtual machines are competing for a limited number of physical cores, which is an unhealthy scheduling scenario. Table \ref{tab:mysql-memcached-cpu-max-min} shows that the best strategy and baseline strategy exhibit smaller differences in CPU utilization, while the worst strategy has a significantly larger difference. This demonstrates that the worst scheduling strategy is suboptimal and requires adjustment.

\section{Design}

This section provides a detailed introduction to the designed scheduler. The scheduler is designed to operate effectively in a heterogeneous core environment, ensuring that it can select appropriate deployment methods for Emulator threads for each virtual machine to maximize their performance.

\subsection{Architectural Design}

\begin{figure}[!htp]
\centering
\includegraphics[width=8cm]{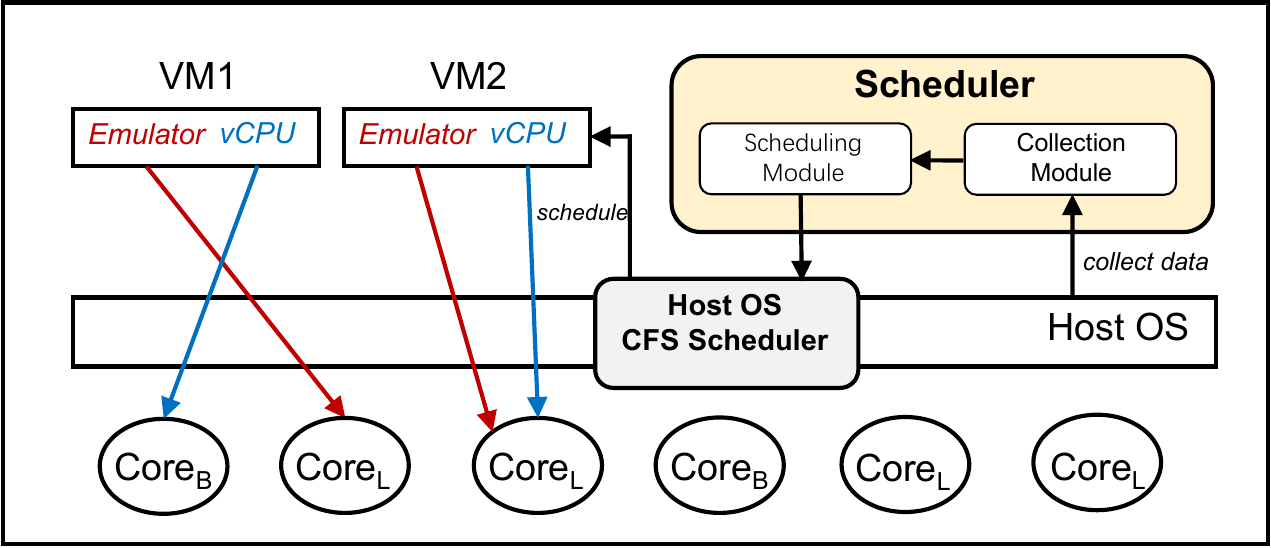} \
\caption{Architectural Overview of the Scheduler}
\label{fig:scheduler-architecture}
\end{figure}

The scheduler is divided into two main modules at the overall architectural level: the collection module and the scheduling module. The holistic architecture is illustrated in Figure~\ref{fig:scheduler-architecture}. The collection module gathers various scheduling metrics and host system status information to gain insights into the operational state of each virtual machine. The scheduling module, employing a finite state automaton design, receives data from the collection module. It dynamically adjusts its state based on the current operational state of the virtual machine and employs various strategies. In distinct states, the scheduling module applies different scheduling policies and ultimately communicates these policies to the host system's thread scheduler, where the final scheduling decisions are executed.

\subsection{Scheduling Strategies}

As outlined in Section~\ref{sec:rundelay-analysis}, the fundamental principle governing scheduling decisions in this scheduler is to balance the scheduling latency between Emulator threads and vCPU threads, aiming to maximize the virtual machine's performance. This scheduler employs the collection module to gather a variety of information, with the scheduling module executing the scheduling decisions.

The central steps of the scheduling strategy involve initially binding all Emulator threads of every virtual machine to all available physical small cores. At this stage, the scheduling latency for Emulator threads is minimal. Subsequently, an attempt is made to progressively reduce the number of bound cores for Emulator threads of each virtual machine. During this reduction process, there is a slight increase in Emulator thread scheduling latency, while the scheduling latency for vCPU threads decreases. The reduction process ceases when the scheduling latency for Emulator threads increases significantly. This approach is repeated when there are fluctuations in the virtual machine's workload or imbalances in CPU utilization within the system.

\subsection{Design of the Collection Module}

This section provides a detailed insight into the design of the collection module. The collection module employs a hierarchical architecture, aggregating information progressively from the bottom up. The collection module consists of two main components: the virtual machine collector and the system collector. The virtual machine collector collects and summarizes information from individual virtual machines through a thread-level collector, while the system collector consolidates data from various lower-level collectors to obtain system-level data. The subsequent subsections will sequentially present the design at each hierarchical level.


The collectors gather data from "proc" system and store the data into a circular buffer (Ring Buffer) in order to get historical information.

\subsection{Scheduling Module Design}

As illustrated in Figure~\ref{fig:scheduler-architecture}, the scheduling module, based on various collected information, makes concrete scheduling decisions and communicates these decisions to the host system's thread scheduler. Emulator threads are scheduled across different physical cores under the combined influence of this scheduling module and the CFS scheduler. The scheduling module's inputs are derived from the information collected by the collection module. The module's output is the optimal physical core scheduling range for Emulator threads in the current context, defining the core binding approach for Emulator threads. The scheduling module employs linux's cgroup API to map its decisions, with the final scheduling being executed by the CFS scheduler.

\subsubsection{Finite State Machine Design}

The core design of the scheduling module is based on a finite state automaton. 



\textbf{Initial State.} In the initial state, the scheduling module conducts initialization tasks. Initialization includes interactions with the collection module to retrieve a list of all currently running virtual machines on the host and capture their current state information. The scheduling module binds all Emulator threads of these virtual machines to all available physical small cores. Subsequently, it transitions to the downscaling state.

\textbf{Downscaling State.} The objective of this state is to balance the scheduling latency of vCPU threads and Emulator threads by reducing the number of physical cores used by Emulator threads, thus maximizing virtual machine performance. Based on the analysis in Section~\ref{sec:rundelay-analysis} and the results presented in Figure~\ref{fig:mysql-memcached-rundelay-cores}, the relationship between Emulator thread scheduling latency and the number of cores to which Emulator threads are bound is approximately inverse. As the number of bound cores increases, Emulator thread scheduling latency decreases rapidly until it reaches a certain point. Beyond this point, the rate of latency reduction decreases noticeably. In the downscaling state, the goal is to reduce the number of cores to this transition point.




The naive approach involves trial and error: reducing the number of cores by one. While the naive approach is straightforward, it is relatively slow. Let $v_1$ be the scheduling latency after the binary core reduction, $n_1$ be the number of cores bound at that point, and $v_2$ be the scheduling latency when Emulator threads are bound to all physical small cores. A parameter $l_1$ is introduced, satisfying $v_1 > v_2,\ l_1 > 1,\ n > n_1$. The definition of a significant increase in scheduling latency is as follows:

\begin{equation}
\frac{v_1 - v_2}{n - n_1} > l_1.\label{eq:rundelay}
\end{equation}

\begin{figure}[!htp]
  \centering
  \includegraphics[width=6cm]{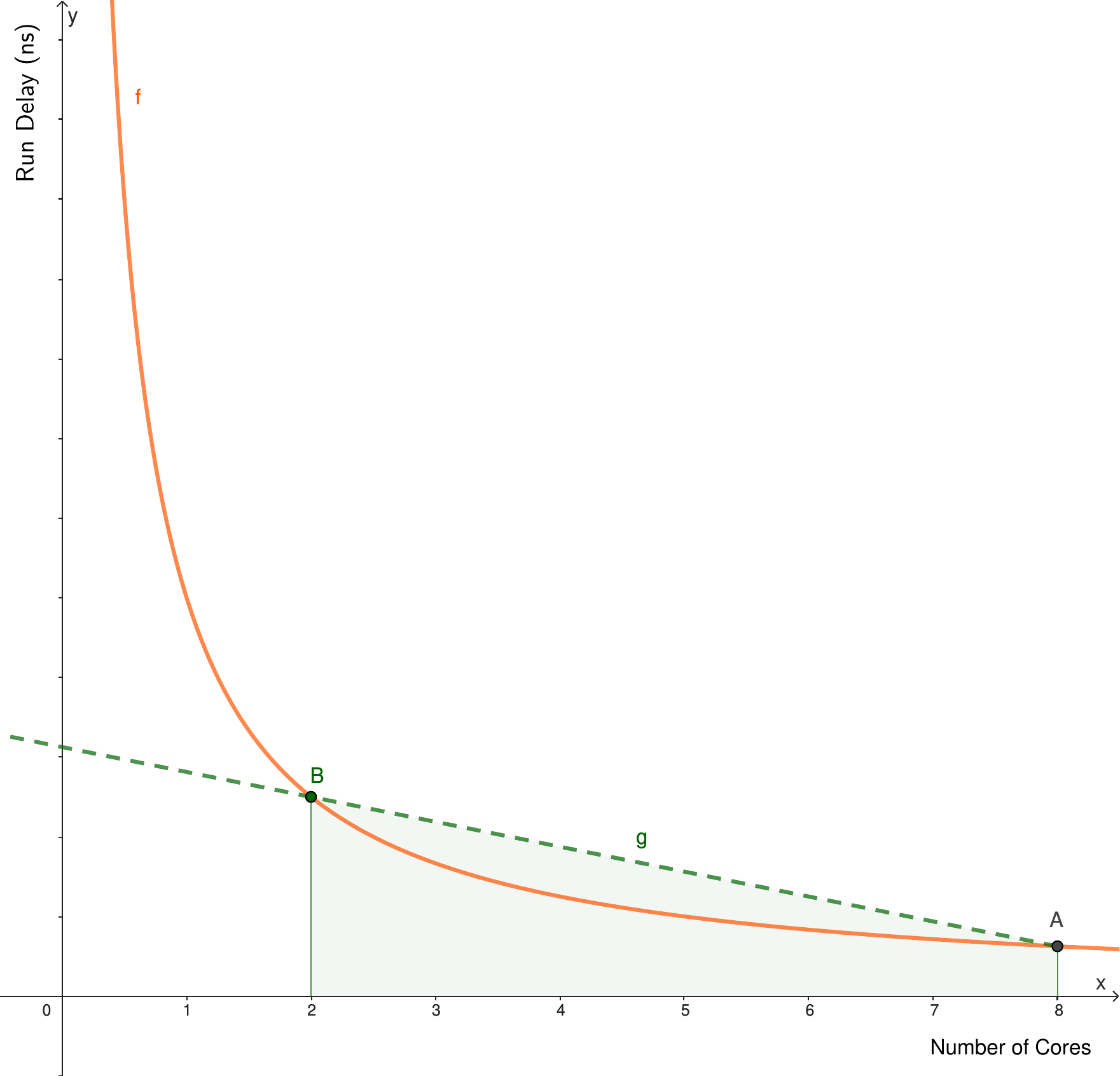}
  \caption{Binary Search on the Scheduling Latency Curve}
  \label{fig:binary-search}
\end{figure}

The figure in Figure~\ref{fig:binary-search} visually represents the binary search process on the scheduling latency curve.

Figure~\ref{fig:binary-search} serves as an illustration of the core concept of binary searching on the scheduling latency curve. The curve $f$ represents the ideal relationship between scheduling latency and the number of cores to which Emulator threads are bound. Let's assume $n = 8$, and point $A$ represents the configuration where Emulator threads are bound to all small cores, with $n = x_A$ and $v_2 = y_A$. The line $g$ has an absolute slope value of $l_1$ and intersects curve $f$ at point B. It is evident that points between $A$ and $B$ do not satisfy Formula~\ref{eq:rundelay}, while points to the left of $B$ do satisfy the formula. Thus, the points in the shaded area in Figure~\ref{fig:binary-search} represent one category of points, and the points outside the shaded area are another category. There is a clear monotonicity, allowing the use of binary searching. 



The time complexity of the binary algorithm is reduced to $O(x\log{n})$, which results in significant speed improvement, especially when $n$ is large.

The downscaling state currently selects one virtual machine for downscaling at a time. The selection strategy is based on the utilization of an Emulator thread in a virtual machine. This strategy aims to maintain the overall balance of CPU usage in the system and make it healthier.

\textbf{Stable State.}
Upon the completion of core reduction for virtual machines, the scheduling module enters a stable state. In this state, the various metrics of both virtual machines and the system remain stable. When the scheduling module is in a stable state, no modifications to core affinity are made. Instead, it collects data from the monitoring module every second to assess whether the entire system remains in a stable state.


The stable state continually monitors three key metrics:
scheduling latency of all virtual machine emulator threads, CPU utilization of vCPU threads and CPU utilization across the system. The method for checking abnormal metric values involves examining whether the magnitude of the change exceeds a certain threshold. 







\textbf{Oscillation State. }The oscillation state serves as a transitional phase between the stable state and the downscaling state. Its purpose is to account for scenarios in which certain metrics are not entirely stable and may exhibit sudden peaks or troughs. To prevent erroneous judgments and unnecessary state transitions, the oscillation state incorporates an upper limit. If specific metrics continuously deviate from their expected values, and the consecutive occurrences exceed the upper limit, the scheduling module enters the downscaling state. Otherwise, the scheduling module returns to the stable state.

When the oscillation state transitions to the downscaling state, only the emulator threads of virtual machines with anomalous data will be rebound to all small cores. Core reduction operations are performed exclusively on these virtual machines, expediting the scheduling process.

\section{Evaluation}

\subsection{Experimental Platform}

\subsubsection{Hardware Configuration}

All experiments were conducted on physical machines. The experimental machine is a dual-socket server, featuring two Huawei Kunpeng 920 processors, with each processor comprising 64 cores, totaling 128 cores, and 256GB of memory. The two processors are divided into 4 Non-Uniform Memory Access (NUMA) nodes. To simulate a heterogeneous core environment, the frequencies of 64 cores in two of the nodes were adjusted to 1.3GHz with a 16MB Last-level Cache (LLC), and they are referred to as "small cores". The other two nodes have 64 cores each, with a frequency of 2.6GHz and a 32MB LLC, referred to as "big cores".

\subsubsection{Software Configuration}

The operating system used is a Linux system with the ARM architecture, running kernel version 5.10.0. A virtualized environment was created using qemu-kvm version 2.8.1.1, and virtual machine management was facilitated through libvirt version 3.2.0. Within the virtual machines, several applications were employed, including MySQL, Memcached, Nginx, Xapian, and PostgreSQL. 



In the experiments, virtual machines were created using the qemu-kvm virtualization technology, and specific configurations were assigned to each virtual machine during the creation process. All KVM virtual machines in the experiment were configured identically, each equipped with 4 virtual processors and 8GB of memory.

\subsection{Evaluation Metrics and Comparison Methods}

All applications used in the experiments are latency-sensitive tasks and have stringent requirements on tail latency. The primary metric used to evaluate the effectiveness of different scheduling strategies is tail latency. In the experiments, benchmarking programs were used to apply static load pressure to the applications. Each test ran for 120 seconds, and the P95 latency of the application was recorded every second. The first and last 10 seconds of data, corresponding to the startup and termination of the application, were excluded. This resulted in a total of 100 data points, which were used to create graphs of tail latency over time. The comparison involved analyzing various aspects of the tail latency graphs, including the average position, maximum position, number of peaks, and other information. To provide a comprehensive evaluation of application performance, the average P95 latency from these 100 data points was used as a single metric for the final performance evaluation.

\subsection{Baseline Scheduling Strategy}

In a conventional homogeneous core environment without different cores, the default placement of Emulator threads aligns with the vCPU threads. For example, a virtual machine configured with 4 cores is generally bound to 4 physical cores, and the Emulator threads are bound to the same set of 4 physical cores. In a heterogeneous multi-core environment, to make optimal use of the respective advantages of big and small cores, cloud service providers or data center administrators may bind the vCPU threads of multiple virtual machines to a combination of physical big and small cores. The Emulator threads are bound to the same physical cores as the vCPU threads. The baseline scheduling strategy involves binding vCPU threads and Emulator threads to the same physical cores. Specifically, on a machine with 4 physical small cores and 4 physical big cores, two virtual machines are co-deployed, with the vCPU threads and Emulator threads of both virtual machines bound to these 8 cores.

\section{Experimental Results}

In the following experiments, two applications, MySQL and Memcached, were co-located in virtual machines. Each virtual machine was configured with 4 cores and 8GB of memory. The vCPU threads of both virtual machines were deployed on 4 small cores and 4 big cores, and the Emulator threads were scheduled using both the baseline scheduling strategy and the scheduler-based scheduling. This section provides a comparison between these two approaches.

\subsubsection{Co-Location of MySQL and Memcached}

\begin{figure}[!htp]
\centering
\subfloat[P95 Latency of MySQL in Co-Location with Memcached\label{fig:mysql-memcached-myql-p95}]{\includegraphics[width=4.2cm]{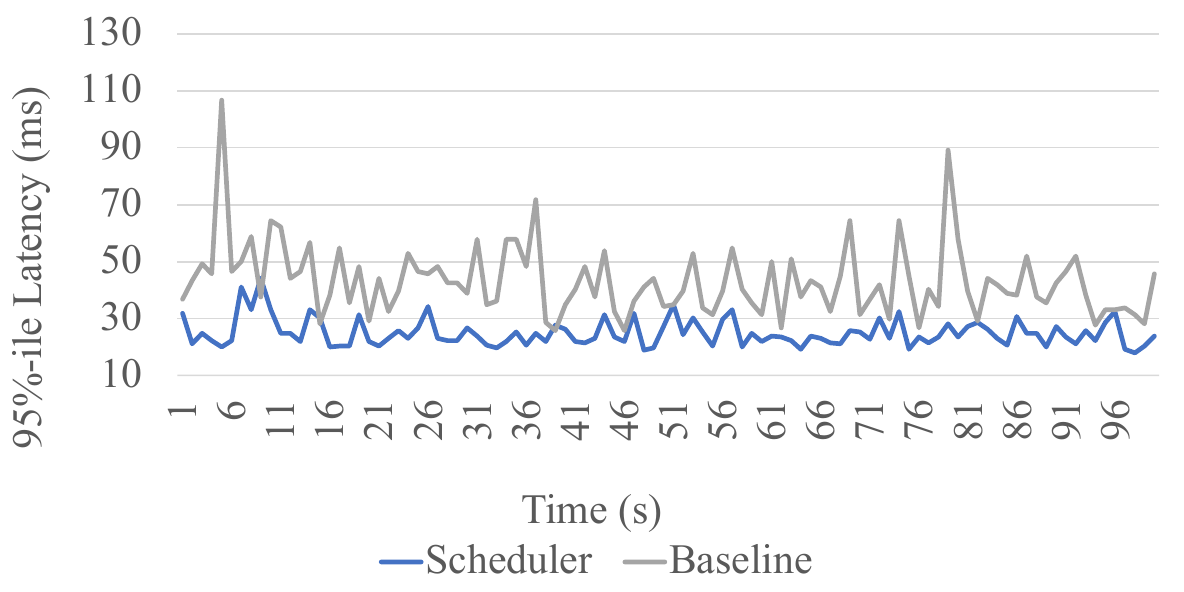}}
\hfil
\subfloat[P95 Latency of Memcached in Co-Location with MySQL\label{fig:mysql-memcached-memcached-p95}]{\includegraphics[width=4.2cm]{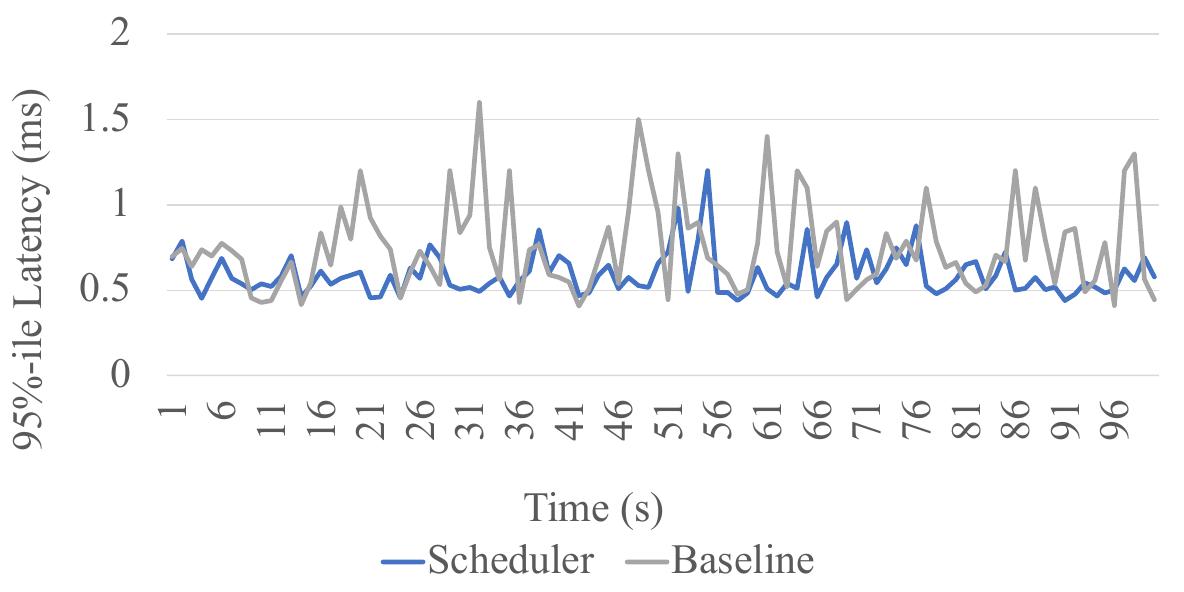}}
\caption{P95 Latency of MySQL and Memcached}
\label{fig:mysql-memcached-p95-and-cores}
\end{figure}



Figures \ref{fig:mysql-memcached-myql-p95} and \ref{fig:mysql-memcached-memcached-p95} display the P95 latency of MySQL and Memcached when co-deployed, respectively. The scheduler improved MySQL's tail latency by 40.7\% and Memcached's tail latency by 20.8\%. With the scheduler, MySQL was bound to 3 small cores, and Memcached was bound to 2 small cores.

\subsubsection{Co-Location of MySQL and Nginx}

\begin{figure}[!htp]
\centering
\subfloat[P95 Latency of MySQL in Co-Location with Nginx\label{fig:mysql-nginx-myql-p95}]{\includegraphics[width=4.2cm]{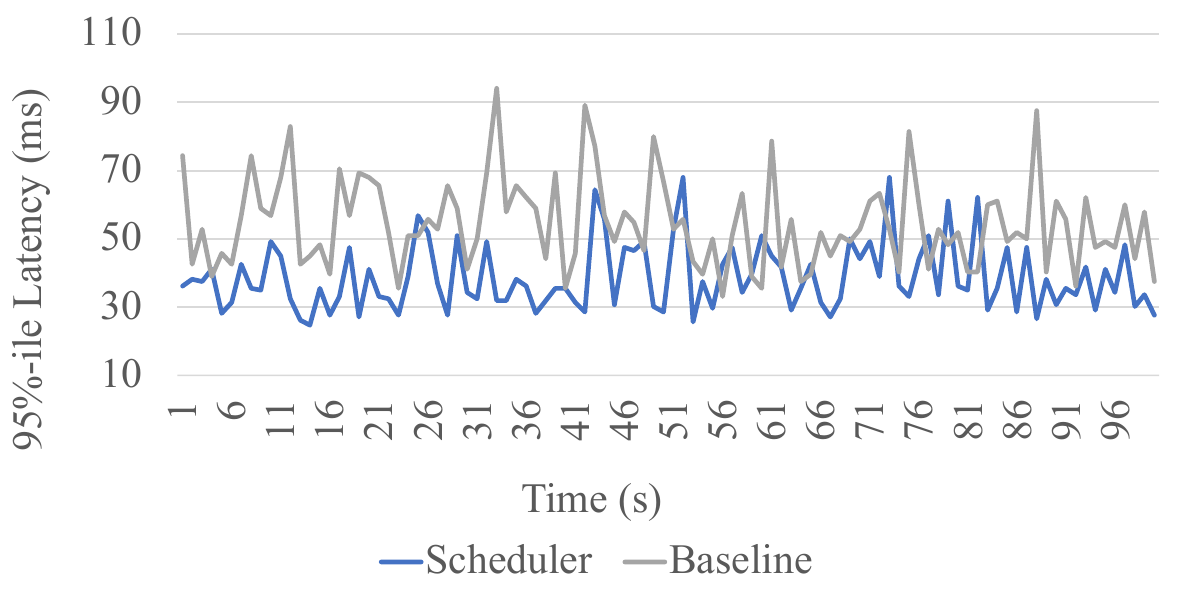}}
\hfil
\subfloat[P95 Latency of Nginx in Co-Location with MySQL\label{fig:mysql-nginx-nginx-p95}]{\includegraphics[width=4.2cm]{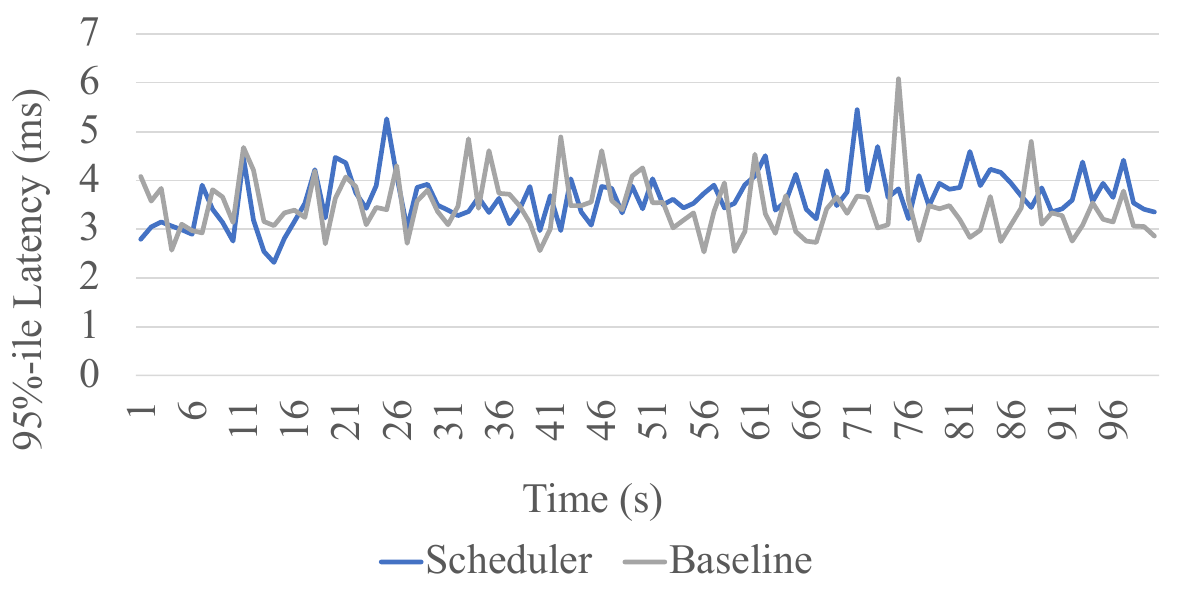}}
\caption{P95 Latency of MySQL and Nginx}
\label{fig:mysql-nginx}
\end{figure}



Figures \ref{fig:mysql-nginx-myql-p95} and \ref{fig:mysql-nginx-nginx-p95} display the P95 latency of MySQL and Nginx when co-deployed, respectively. The scheduler improved MySQL's tail latency by 34.4\%, while Nginx's tail latency remained nearly the same as the baseline. With the scheduler, MySQL was bound to 1 small core, and Nginx was bound to 1 small core.

\subsubsection{Co-Location of MySQL and Xapian}

\begin{figure}[!htp]
\centering
\subfloat[P95 Latency of MySQL in Co-Location with Xapian\label{fig:mysql-xapian-myql-p95}]{\includegraphics[width=4.2cm]{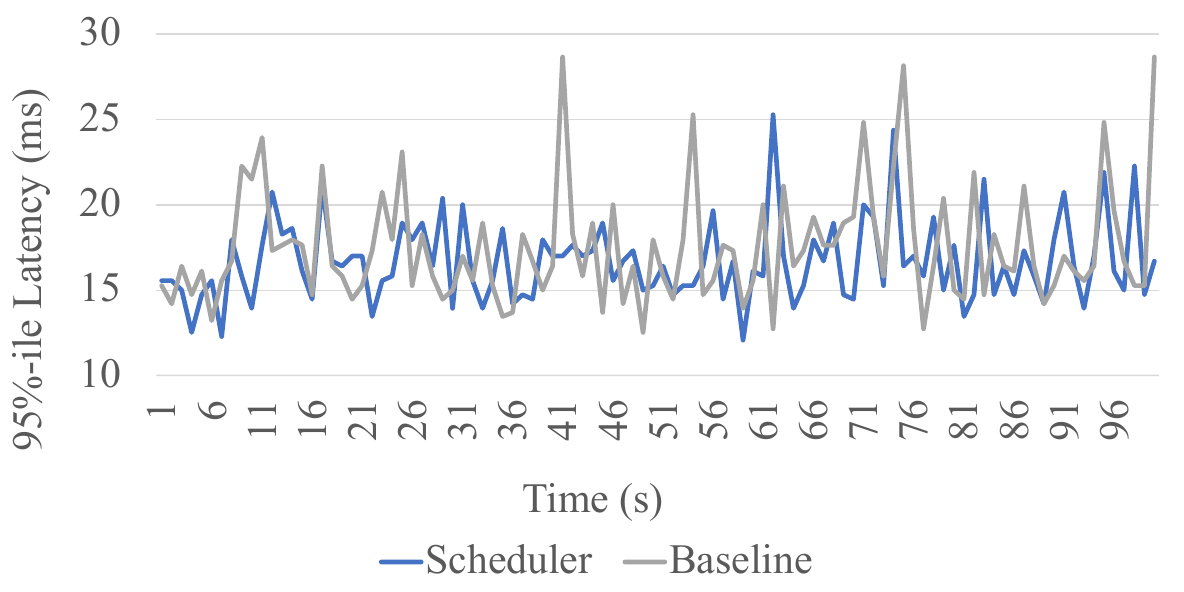}}
\hfil
\subfloat[P95 Latency of Xapian in Co-Location with MySQL\label{fig:mysql-xapian-xapian-p95}]{\includegraphics[width=4.2cm]{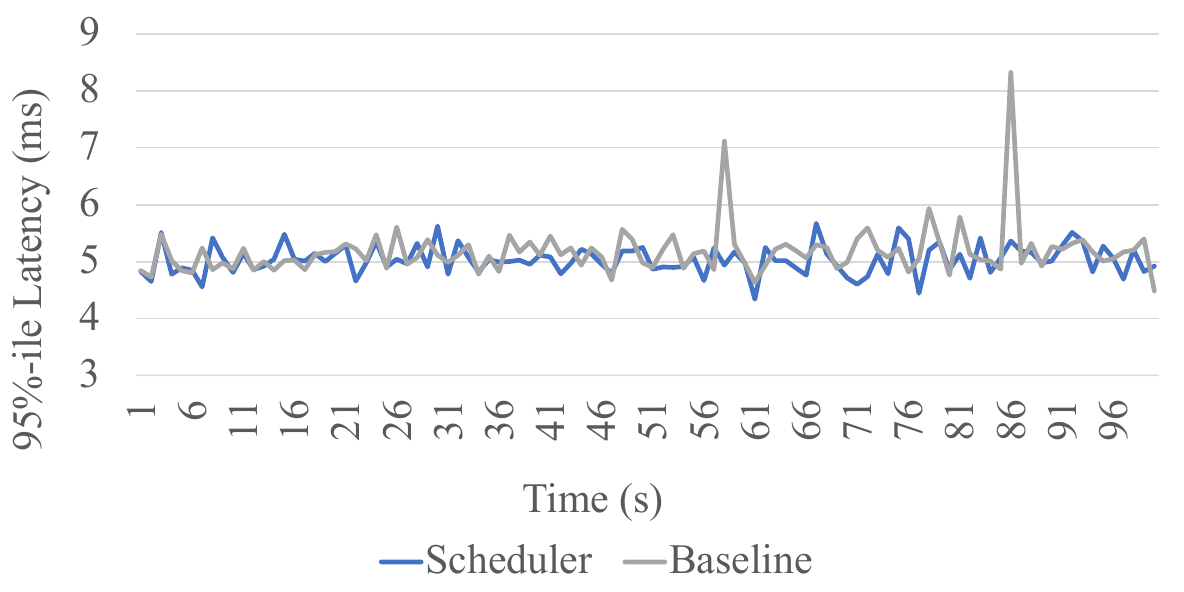}}
\caption{P95 Latency of MySQL and Xapian}
\label{fig:mysql-xapian}
\end{figure}



Figures \ref{fig:mysql-xapian-myql-p95} and \ref{fig:mysql-xapian-xapian-p95} display the P95 latency of MySQL and Xapian when co-deployed, respectively. The scheduler improved MySQL's tail latency by 4.13\% and Xapian's tail latency by 7.86\%. With the scheduler, MySQL was bound to 3 small cores, and Xapian was bound to 4 small cores.

\subsubsection{Co-Location of PostgreSQL and Memcached}

\begin{figure}[!htp]
\centering
\subfloat[P95 Latency of PostgreSQL in Co-Location with Memcached\label{fig:pg-memcached-pg-p95}]{\includegraphics[width=4.2cm]{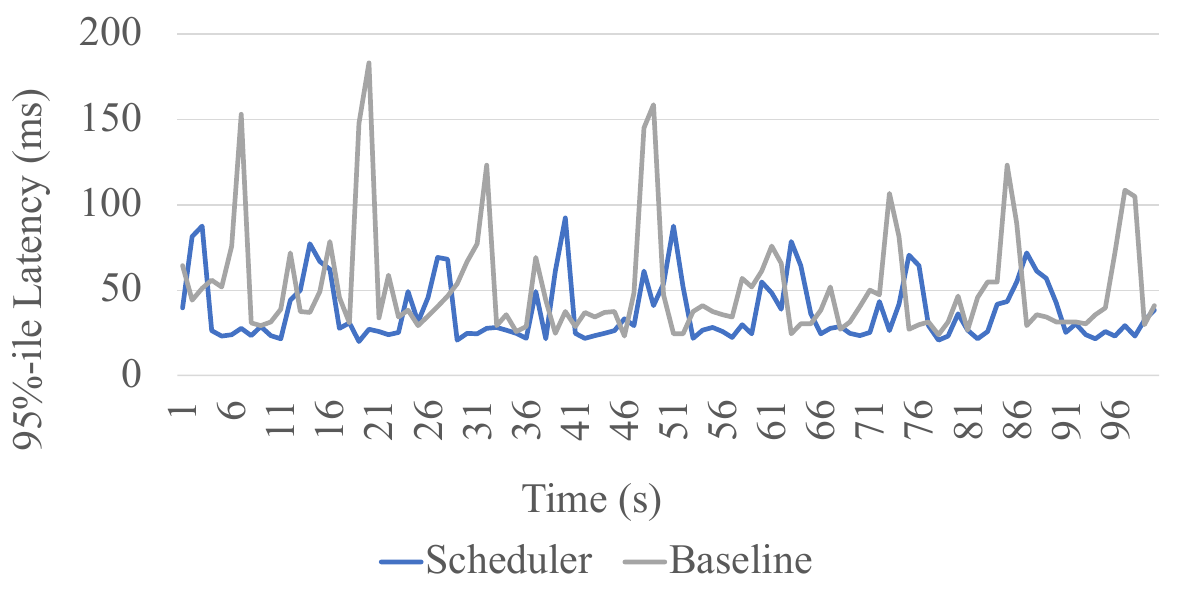}}
\hfil
\subfloat[P95 Latency of Memcached in Co-Location with PostgreSQL\label{fig:pg-memcached-memcached-p95}]{\includegraphics[width=4.2cm]{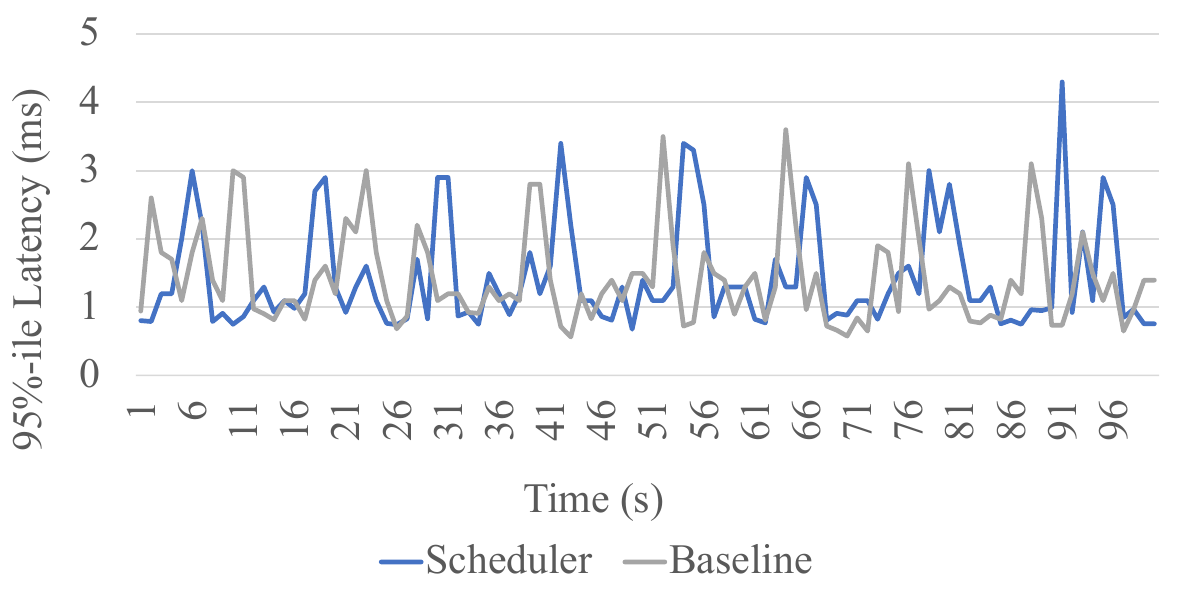}}
\caption{P95 Latency of PostgreSQL and Memcached}
\label{fig:pg-memcached}
\end{figure}



Figures \ref{fig:pg-memcached-pg-p95} and \ref{fig:pg-memcached-memcached-p95} show the P95 latency of PostgreSQL and Memcached when co-deployed, respectively. The scheduler improved PostgreSQL's tail latency by 27.0\%, while the results for Memcached are almost identical to the baseline. With the scheduler, PostgreSQL was bound to 4 small cores, and Memcached was bound to 4 small cores.

\subsubsection{Co-Location of PostgreSQL and Nginx}

\begin{figure}[!htp]
\centering
\subfloat[P95 Latency of PostgreSQL in Co-Location with Nginx\label{fig:pg-nginx-pg-p95}]{\includegraphics[width=4.2cm]{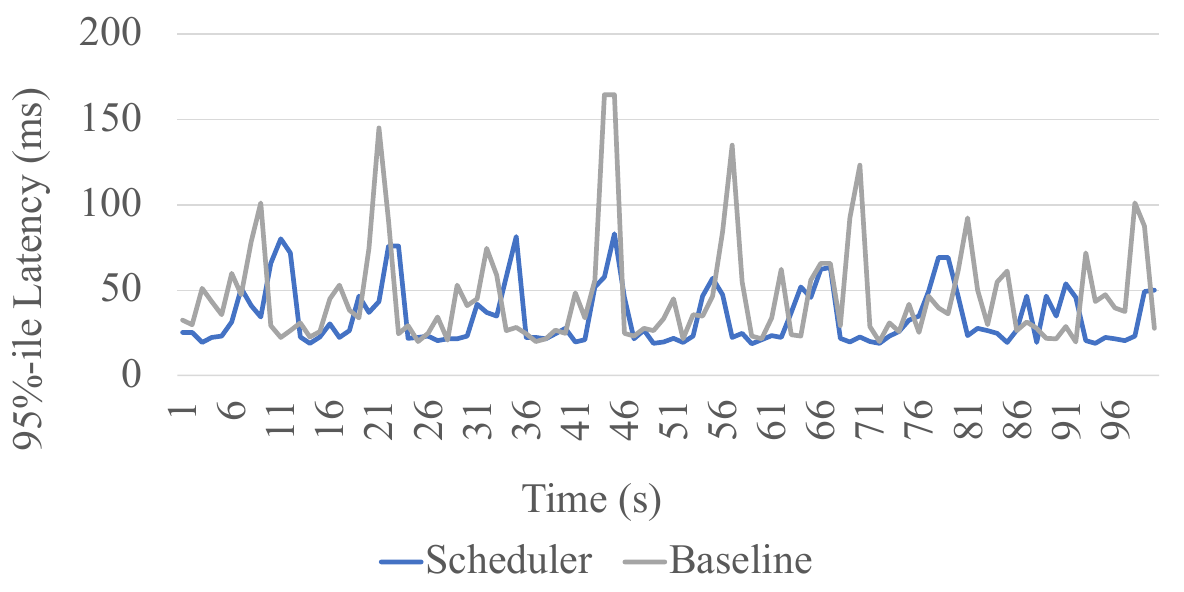}}
\hfil
\subfloat[P95 Latency of Nginx in Co-Location with PostgreSQL\label{fig:pg-nginx-nginx-p95}]{\includegraphics[width=4.2cm]{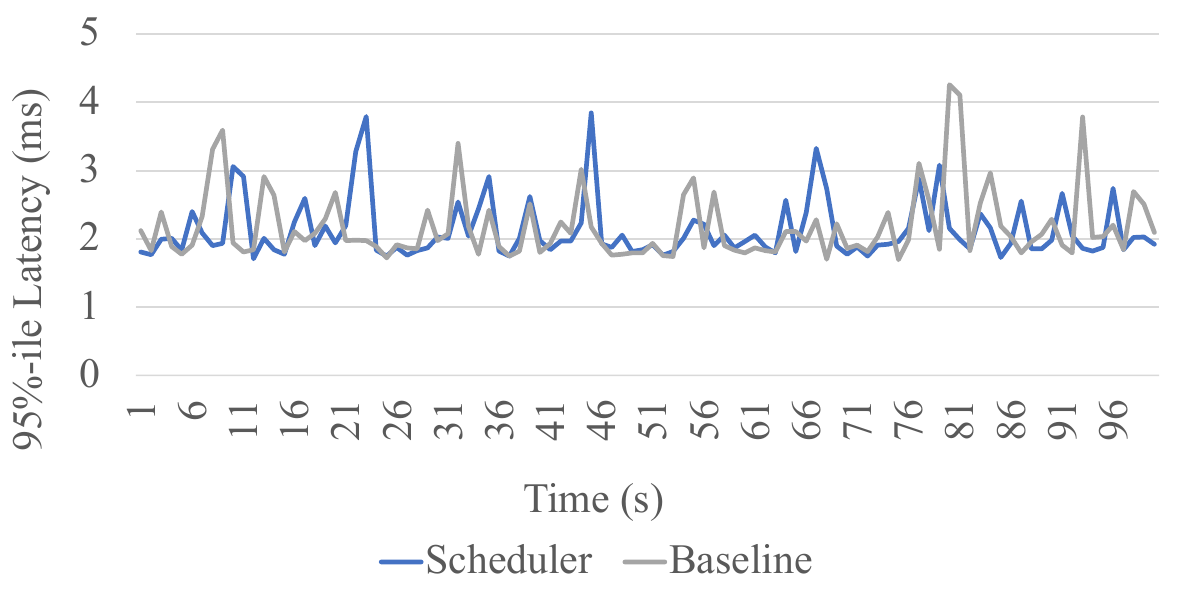}}
\caption{P95 Latency of PostgreSQL and Nginx}
\label{fig:pg-nginx}
\end{figure}



Figures \ref{fig:pg-nginx-pg-p95} and \ref{fig:pg-nginx-nginx-p95} show the P95 latency of PostgreSQL and Nginx when co-deployed, respectively. The scheduler improved PostgreSQL's tail latency by 25.4\% and improved Nginx's tail latency by 2.47\%. With the scheduler, PostgreSQL was bound to 3 small cores, and Nginx was bound to 4 small cores.

\subsubsection{Co-Location of PostgreSQL and Xapian}

\begin{figure}[!htp]
\centering
\subfloat[P95 Latency of PostgreSQL in Co-Location with Xapian\label{fig:pg-xapian-myql-p95}]{\includegraphics[width=4.2cm]{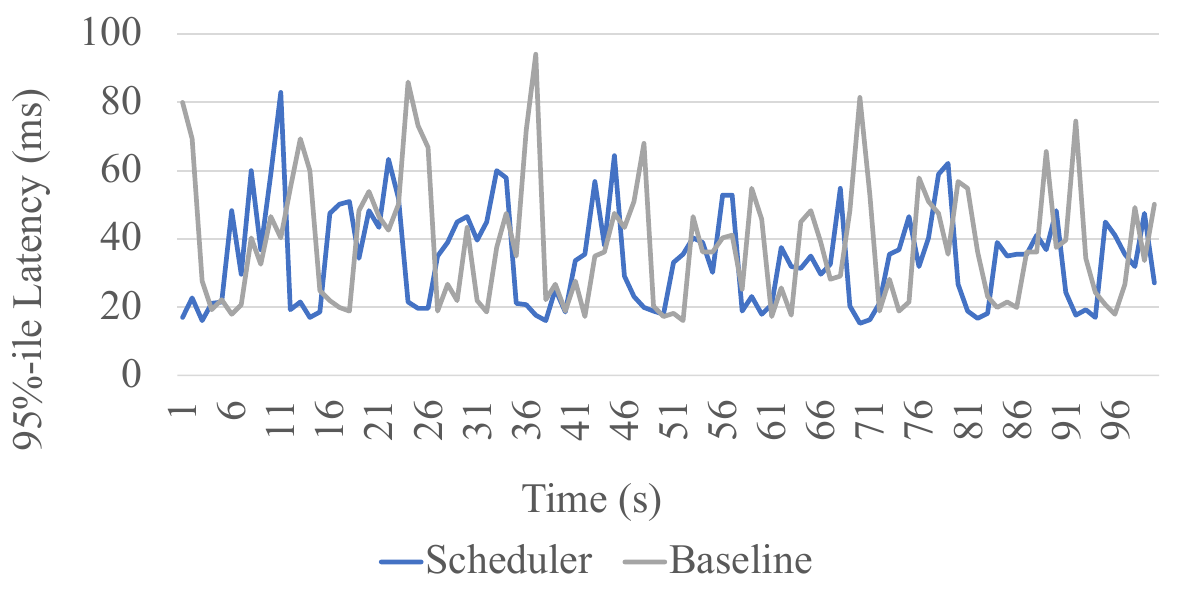}}
\hfil
\subfloat[P95 Latency of Xapian in Co-Location with PostgreSQL\label{fig:pg-xapian-xapian-p95}]{\includegraphics[width=4.2cm]{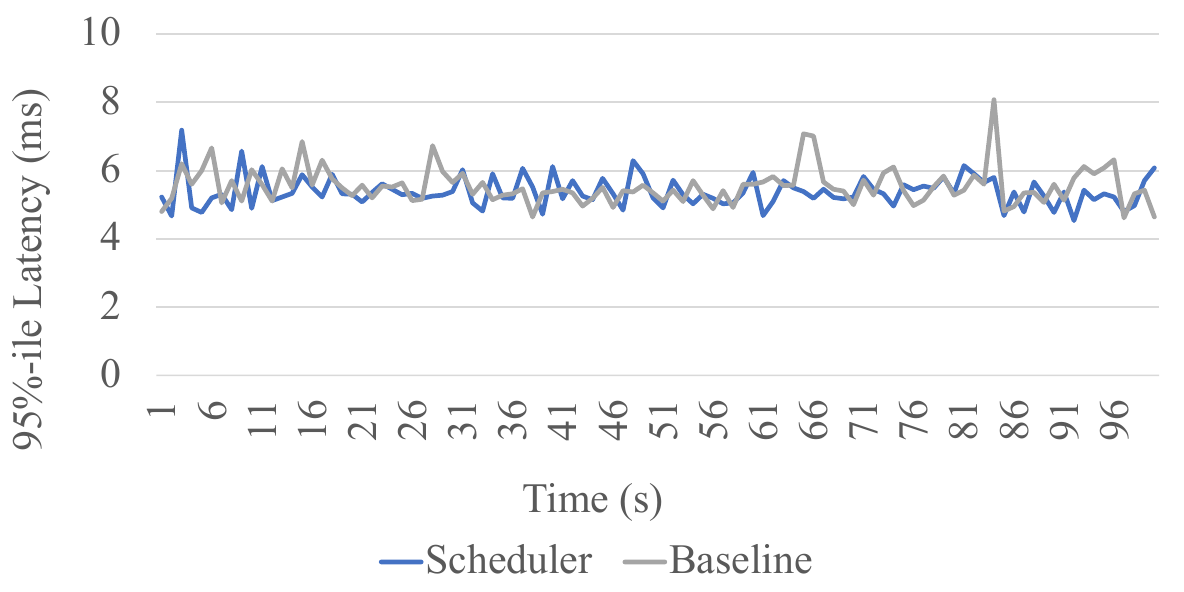}}
\caption{P95 Latency of PostgreSQL and Xapian}
\label{fig:pg-xapian}
\end{figure}



Figures \ref{fig:pg-xapian-myql-p95} and \ref{fig:pg-xapian-xapian-p95} show the P95 latency of PostgreSQL and Xapian when co-deployed, respectively. The scheduler improved PostgreSQL's tail latency by 11.5\% and improved Xapian's tail latency by 3.25\%. With the scheduler, PostgreSQL was bound to 4 small cores, and Xapian was bound to 4 small cores.

\subsubsection{Co-Location of MySQL and PostgreSQL}

\begin{figure}[!htp]
\centering
\subfloat[P95 Latency of MySQL in Co-Location with PostgreSQL\label{fig:mysql-pg-myql-p95}]{\includegraphics[width=4.2cm]{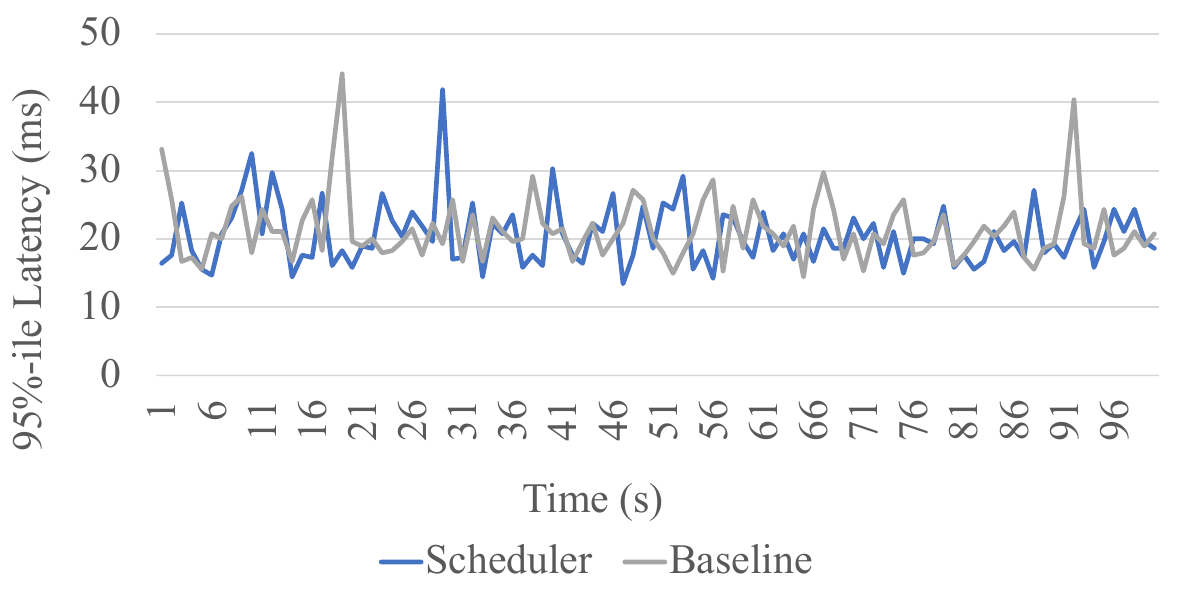}}
\hfil
\subfloat[P95 Latency of PostgreSQL in Co-Location with MySQL\label{fig:mysql-pg-pg-p95}]{\includegraphics[width=4.2cm]{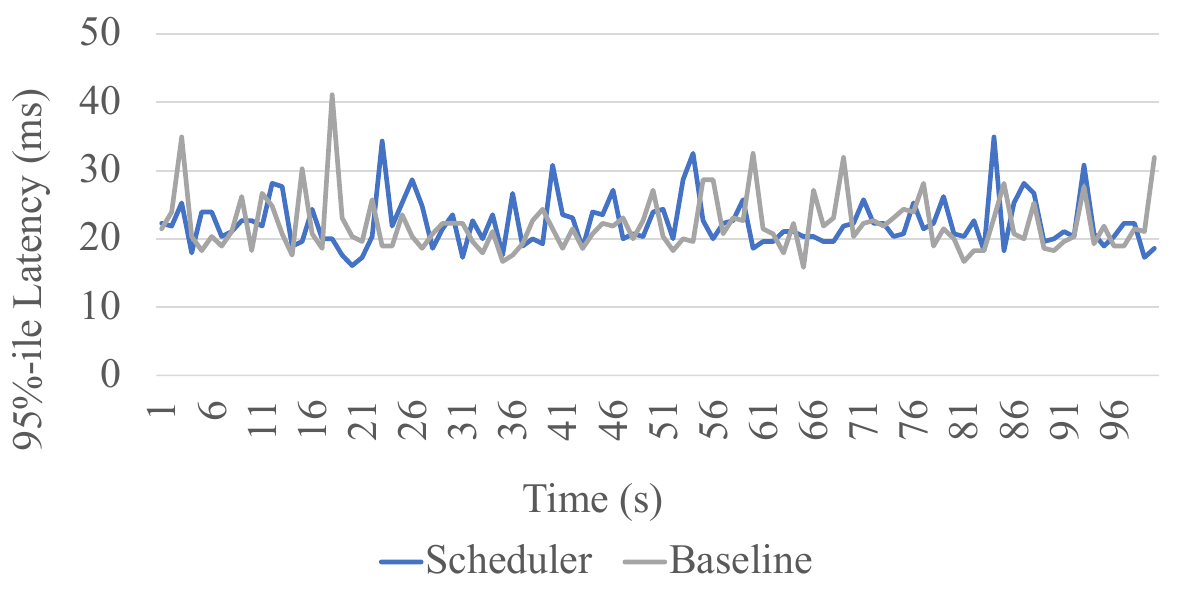}}
\caption{P95 Latency of MySQL and PostgreSQL}
\label{fig:mysql-pg}
\end{figure}



Figures \ref{fig:mysql-pg-myql-p95} and \ref{fig:mysql-pg-pg-p95} show the P95 latency of MySQL and PostgreSQL when co-deployed, respectively. The scheduler improved MySQL's tail latency by 4.18\%, while PostgreSQL's tail latency remained nearly the same as the baseline. With the scheduler, MySQL was bound to 2 small cores, and PostgreSQL was bound to 4 small cores.

\section{Conclusion}

In this study, we delve into the scheduling strategy of Emulator threads within virtual machine processes in a heterogeneous computing environment. We investigate the utilization of Emulator threads, the optimal number of core bindings, and the impact of their scheduling strategy on virtual machine performance. Based on a scheduling latency metric, a scheduler has been designed to dynamically adjust the core bindings of Emulator threads in response to the virtual machine's state. Experimental validation demonstrates that the scheduler effectively enhances virtual machine performance, with the maximum observed performance improvement being 40.7\%.

\bibliographystyle{IEEEtran}
\bibliography{main}

\end{document}